\documentclass[twocolumn,12pt]{aastex6}
\usepackage{natbib}
\usepackage{verbatim}

\begin{document}

\title{The Solar Wind Environment in Time}

\author{Quentin Pognan\altaffilmark{1}, Cecilia
  Garraffo\altaffilmark{1}, Ofer Cohen\altaffilmark{2,1}, Jeremy J. Drake\altaffilmark{1}}

\altaffiltext{1}{Harvard-Smithsonian Center for Astrophysics, 60 Garden St. Cambridge, MA 02138}
\altaffiltext{2}{Lowell Center for Space Science and Technology, University of Massachusetts, Lowell, MA 01854, USA}


\begin{abstract}

We use magnetograms of 8 solar
analogues of ages 30~Myr to 3.6~Gyr obtained from Zeeman Doppler Imaging (ZDI) and taken from the literature, together with two solar magnetograms, to drive magnetohydrodynamical (MHD) wind simulations and construct an evolutionary scenario of the solar wind environment and its angular momentum loss rate. With observed magnetograms of the radial field strength as the only variant in the wind model, we find that power law model fitted to the derived angular momentum loss rate against time, $t$, results in a spin down relation $\Omega\propto t^{-0.51}$, 
for angular speed $\Omega$, which is remarkably consistent with the well-established Skumanich law $\Omega\propto t^{-0.5}$. We use the model wind conditions to estimate the magnetospheric standoff distances for an Earth-like test planet situated at 1~AU for each of the stellar cases, and to obtain trends of minimum and maximum wind ram pressure and average ram pressure in the solar system through time.  The wind ram pressure declines with time as $\overline{P_{ram}}\propto t^{2/3}$, amounting to a factor of 50 or so over the present lifetime of the solar system.
 
\end{abstract}


\section{INTRODUCTION}
\label{sec:Intro}

As with all stars, the Sun has undergone significant changes during its
lifetime.  In addition to the substantial structural changes from its pre-main sequence phase through to evolution on the main-sequence \citep[e.g.][]{Sagan.Mullen:72,Guzik.Cox:95,Sackmann.Boothroyd:03}, its rotation and associated magnetic
activity have also changed quite radically \citep[e.g.][]{Gudel:07,Guinan.Engle:09}.
Understanding the Sun's past is of fundamental importance for understanding the solar system as a whole. Changes in radiative energy, magnetic field, the solar wind and more
transient phenomena such as flares and coronal mass ejections have inevitably played a major role in
planetary evolution and the appearance of life on
Earth. Realisation has been growing that the Sun's magnetic activity is particularly important.  It is the source of UV--X-ray radiation that would likely have driven significant mass loss from planetary envelopes early in solar system history \citep{Owen.Wu:16}, and of the solar wind thought to be the culprit behind the disappearance of water from Mars during the Noachian period \citep[e.g.][]{Terada.etal:09}.  The solar wind is also the 
driving force behind the heliospheric bubble that has protected the solar system from
potentially harmful cosmic rays, both today and during the epoch of emergent life on Earth \citep{Cohen.etal:12}. 

The magnetic activity of a star is dependent on
its magnetic dynamo, which in turn harbours a strong relation with the rotation
period of the star \citep[e.g.][]{Skumanich:72,Pallavicini.etal:81,Wright.Drake:11}. In accordance with main sequence development, the Sun is thought to have been rotating much faster in its younger days
by up to 10 times or so \citep{Guinan.Engle:09}, which would have
induced a much more vigorous dynamo and stronger magnetic activity
\citep[e.g.][]{Gudel:07}. This implies a significantly
different solar environment in the past than the present day one. 

Magnetized stellar winds cause mass and angular momentum
loss when they break away from the star's magnetic field in a process called magnetic braking \citep{Schatzman:62,Weber.Davis:67,Mestel:68,Mestel.Spruit:87}. It is theorised that these winds are driven by magnetic processes ultimately powered by the stellar magnetic dynamo. Although the exact physics involved has not been unequivocally identified, dissipation of Alfv\'en wave pressure \citep{Belcher:71} is gaining widespread acceptance as a dominant mechanism \citep[e.g.][]{Suzuki:06,Cranmer.van_Ballegooijen:05,van_Ballegooijen.etal:11}. 

Since the dynamo has an intrinsic link to the star's
rotation period, it is also related to the age of the star. It is known that stellar 
rotation rates, $\Omega$, eventually converge to the empirically observed Skumanich
Law, $\Omega \sim t^{-1/2}$ \citep{Skumanich:72}. However the
relation is not very well understood for the early stages of stellar
evolution. Recent surveys of young open
clusters \citep[e.g.][]{Meibom.etal:11} have found a wide range of
rotation periods for younger stars. Thus it is a safe assumption that the
Sun must have had very different angular braking and mass loss rates during
its earlier lifetime. 


In order to understand how the Sun behaved in past stages of its
life we can look to similar but younger solar analogue stars
\citep{Gudel:07,Guinan.Engle:09,Ribas.etal:05,Ribas.etal:10} which can provide insight into the 
solar environment at earlier times. With some understanding of the powering of the present day solar wind by the surface magnetic field, 
observations of the magnetic fields of solar analogues can also be used to attempt to constrain the Sun's
stellar wind environment at various points
in its life \citep[see, e.g.,][for a proof of concept example]{Sterenborg.etal:11}. Current space weather models \cite[e.g.][]{Cohen.etal:08} utilize detailed solar magnetograms obtained using
Michelson Doppler Imaging (MDI) \citep{Scherrer.etal:95} as the lower magnetic field boundary condition. With recent developments in instrumentation
and observational techniques, one of the most powerful methods for acquiring
magnetic field data on other late-type stars is Zeeman Doppler Imaging 
\citep[ZDI; see, e.g.,][]{Brown.etal:91,Donati.Brown:97,Hussain.etal:09} from which
large-scale magnetograms of stellar surfaces can be
retrieved. 

In this study, we look at solar analogues of various ages as proxies for the
Sun in its youth. We present the results of numerical simulations in
the solar coronal regime for 8 solar-type stars with ages ranging from
30~Myr to 3.6~Gyr. From the models we extract the rates for mass
and angular momentum losses of each star, as well as the range of wind
speeds and associated densities. The wind speeds and densities are
extrapolated from 1~Au up to 100~AU to examine the ram pressures
at distances comparable to those of solar system planets from the Sun. Notably the 1~Au distance is used to calculate magnetospheric standoff radii to examine the consequent implications for the terrestrial magnetosphere.
We also make use of
two solar magnetograms, one with high resolution and the other with
resolution closer to that of the ZDI maps, in order to test the
effects of limited magnetogram resolution obtained from the ZDI approach. 

We begin by detailing the numerical
magnetohydrodynamics (MHD)
simulations in Section~\ref{sec:Methods}. We then present the results of the modeling in
Section~\ref{sec:Results} and discuss the main findings in Section~\ref{sec:Discussion}. We finish
with Section~\ref{sec:Conclusions}, where we conclude that the use of ZDI magnetograms of solar analogues is a valid method for studying the Sun's past and should be continued with the advent of future data. \\
\\

\begin{figure*}
\center
\includegraphics[trim = 0.3in 0.3in
  0.5in 3.5in,clip, width = 0.4\textwidth]{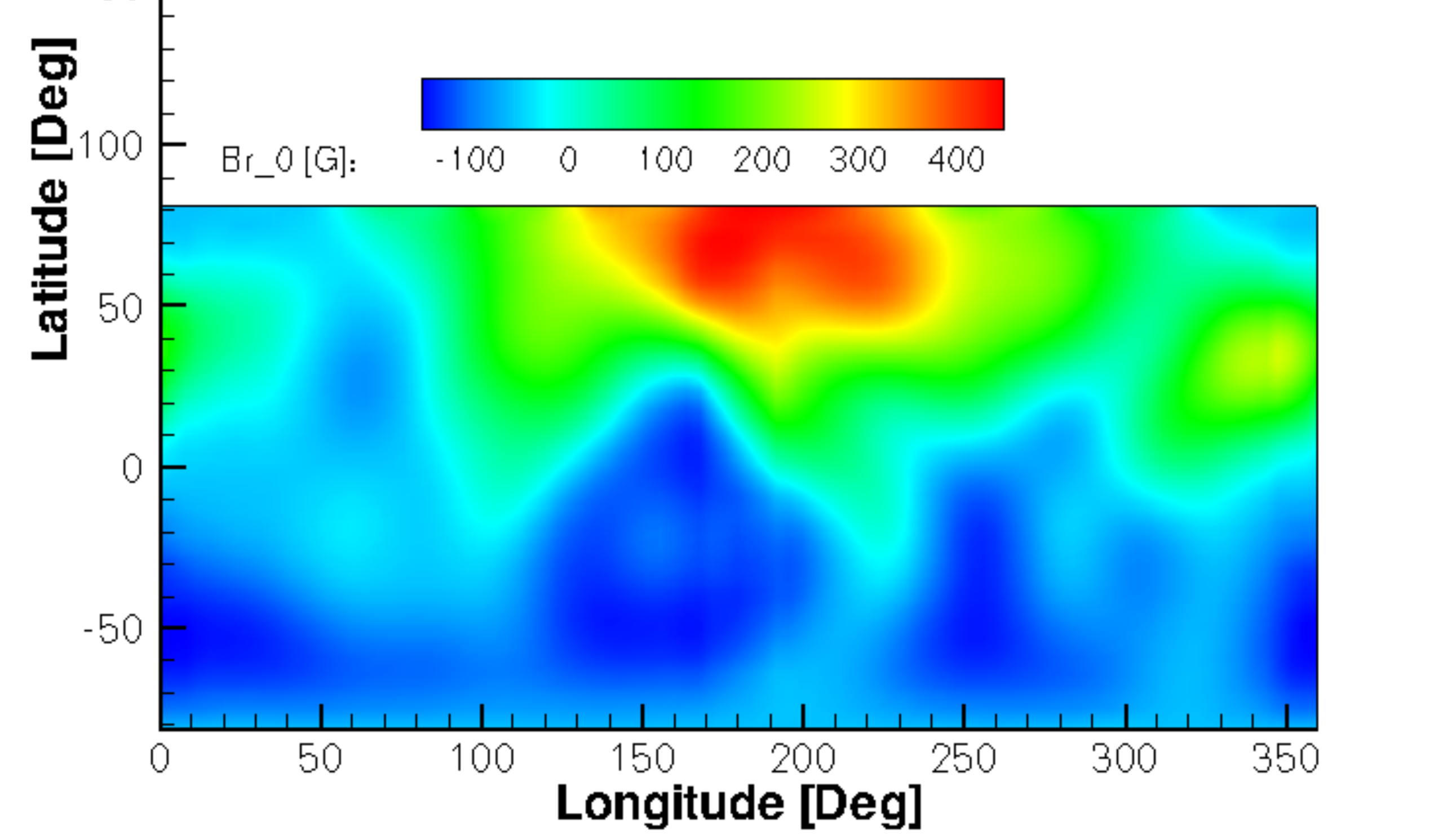} 
\includegraphics[trim = 0.3in 0.3in
  0.5in 3.5in,clip, width = 0.4\textwidth]{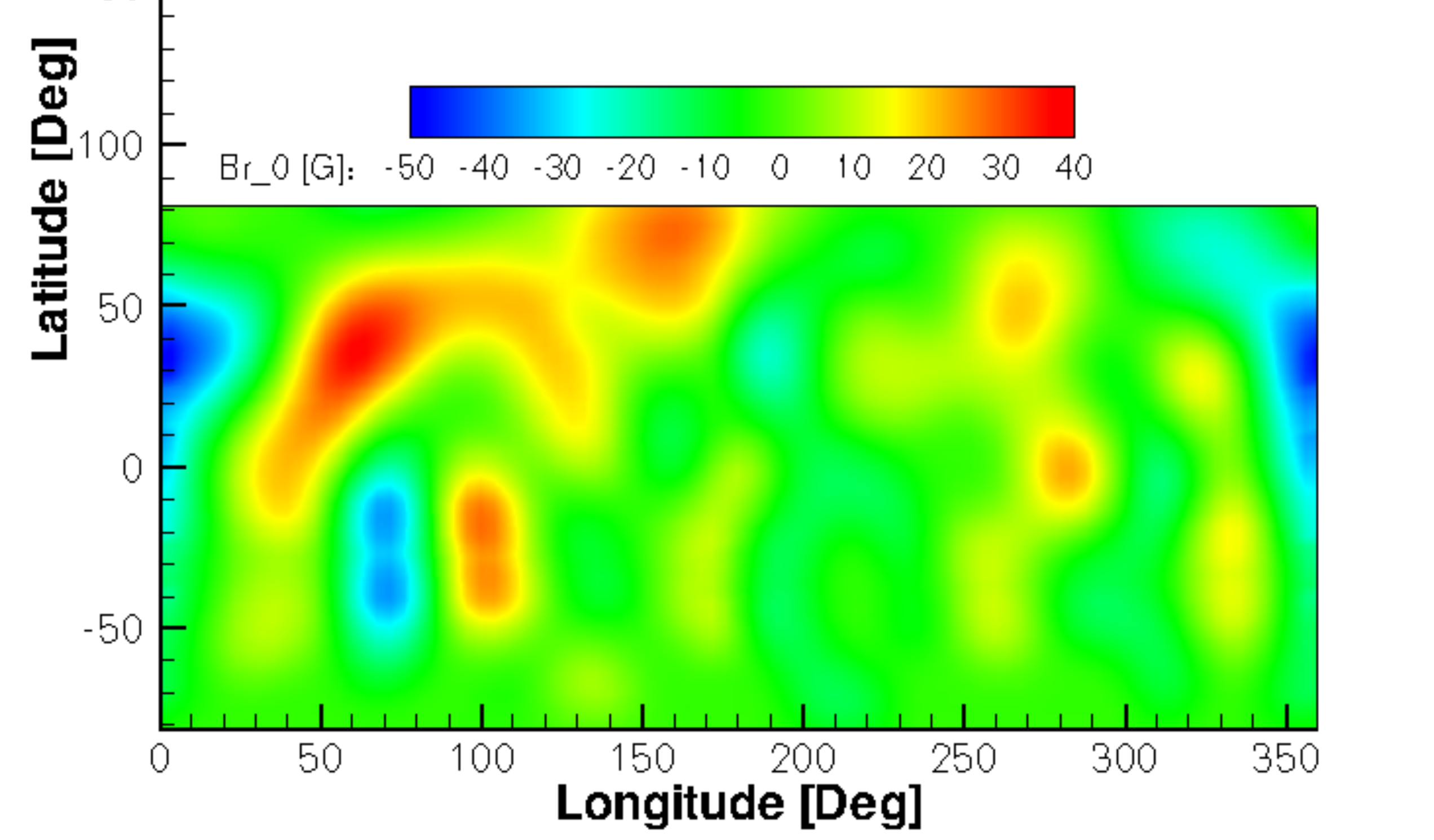}\\ 
\includegraphics[trim = 0.3in 0.27in
  0.5in 3.5in,clip, width = 0.4\textwidth]{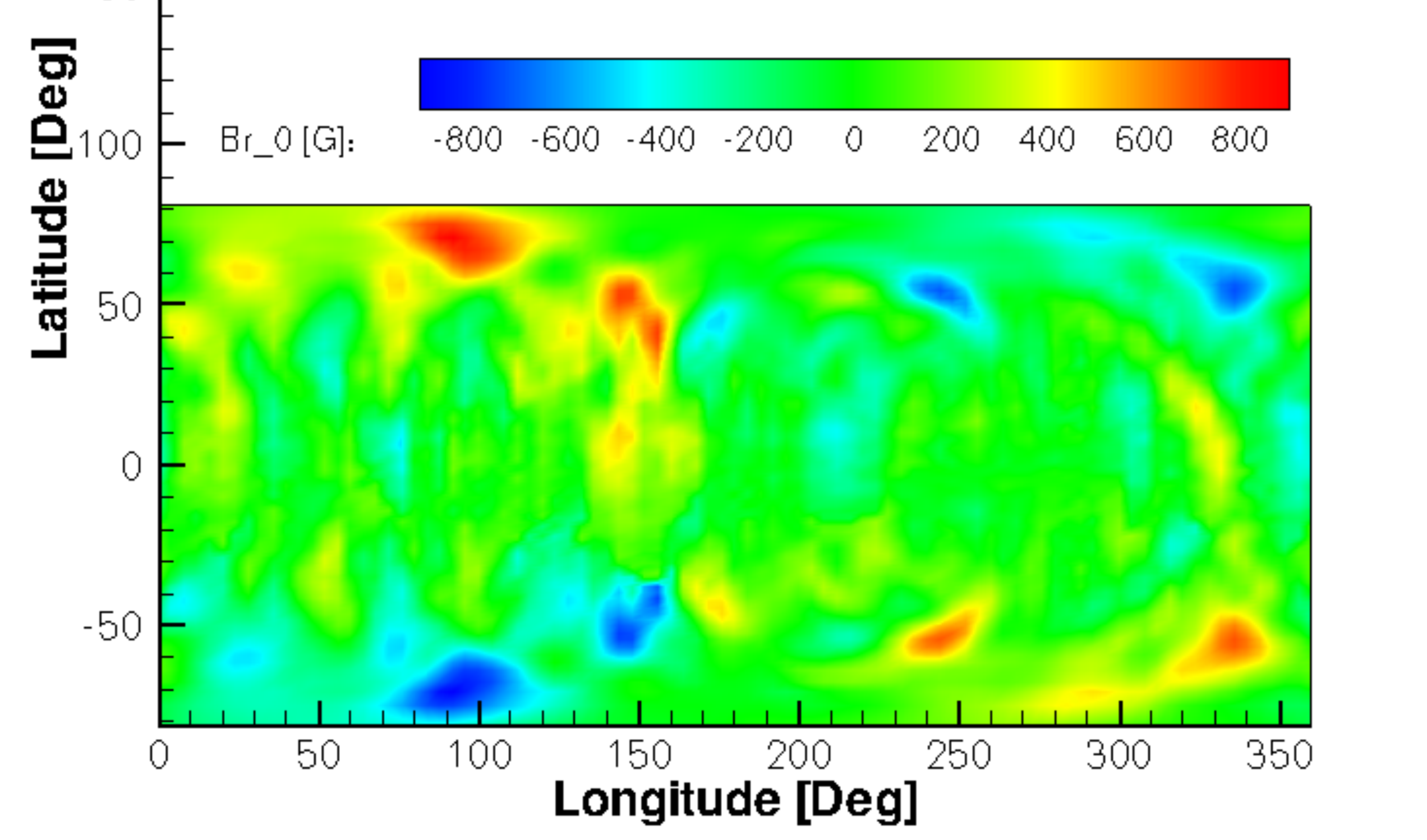} 
\includegraphics[trim = 0.3in 0.3in
  0.5in 3.5in,clip, width = 0.4\textwidth]{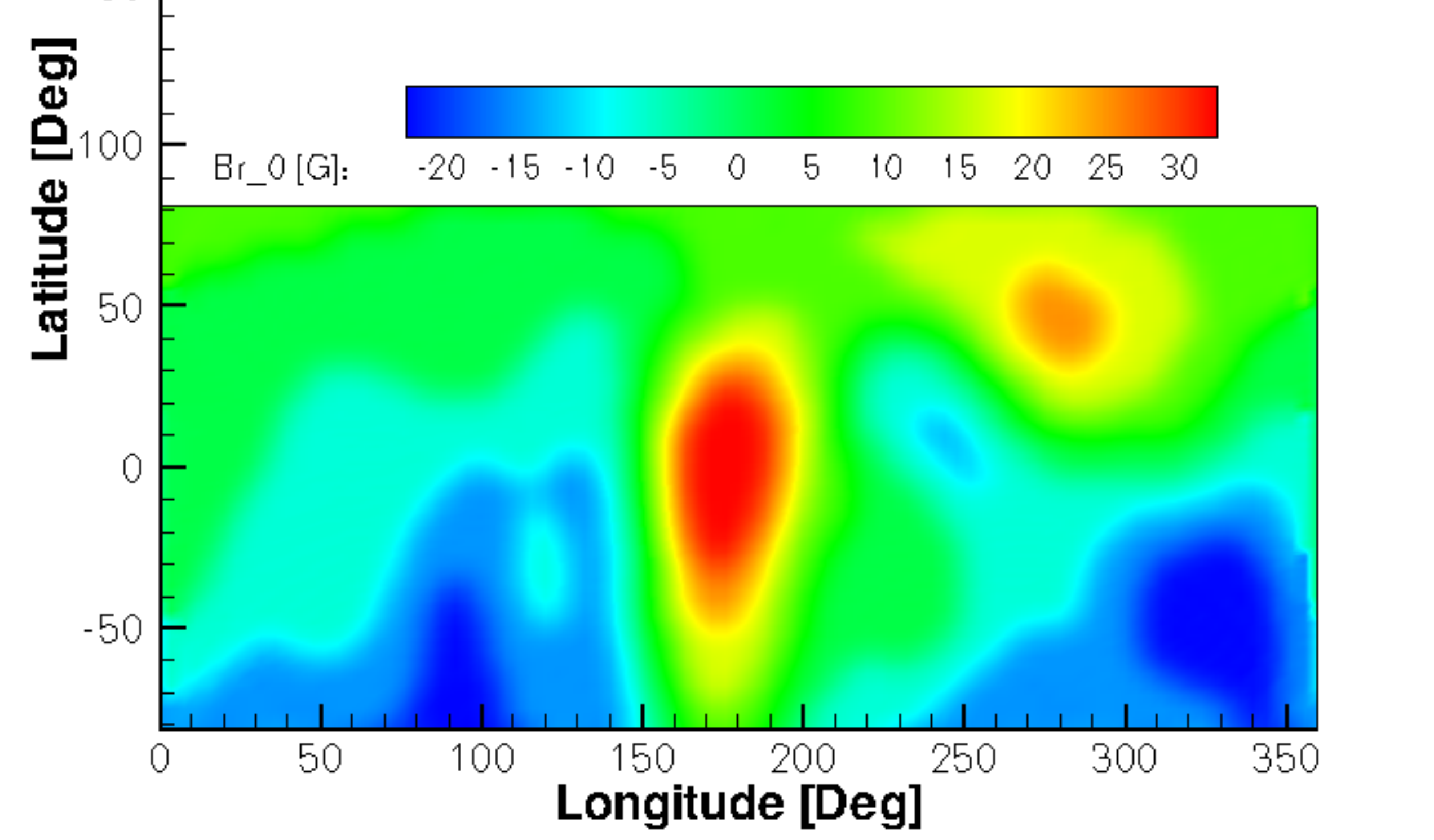}\\ 
\includegraphics[trim = 0.3in 0.3in
  0.5in 3.5in,clip, width = 0.4\textwidth]{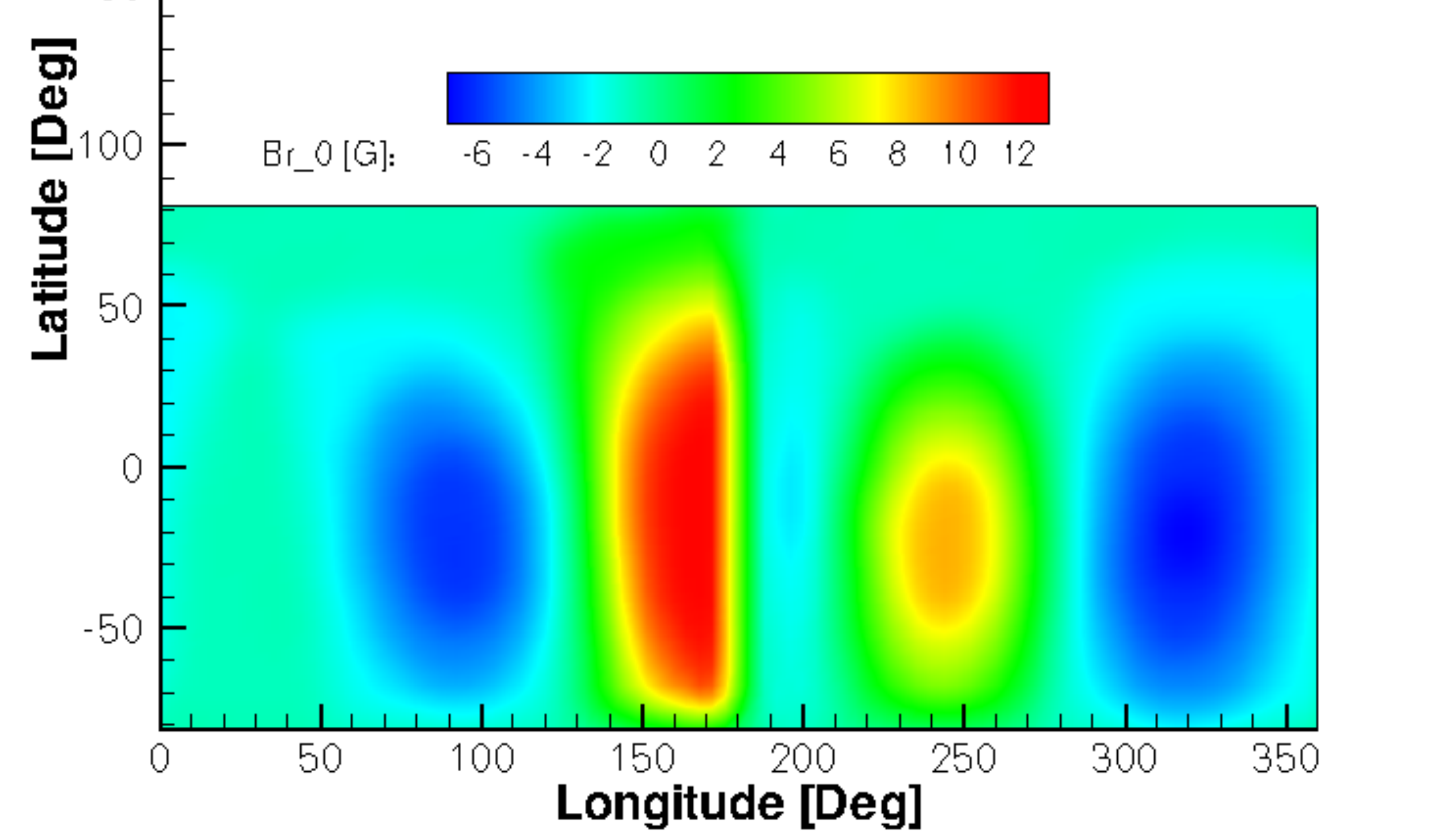} 
\includegraphics[trim = 0.3in 0.3in
  0.5in 3.5in,clip, width = 0.4\textwidth]{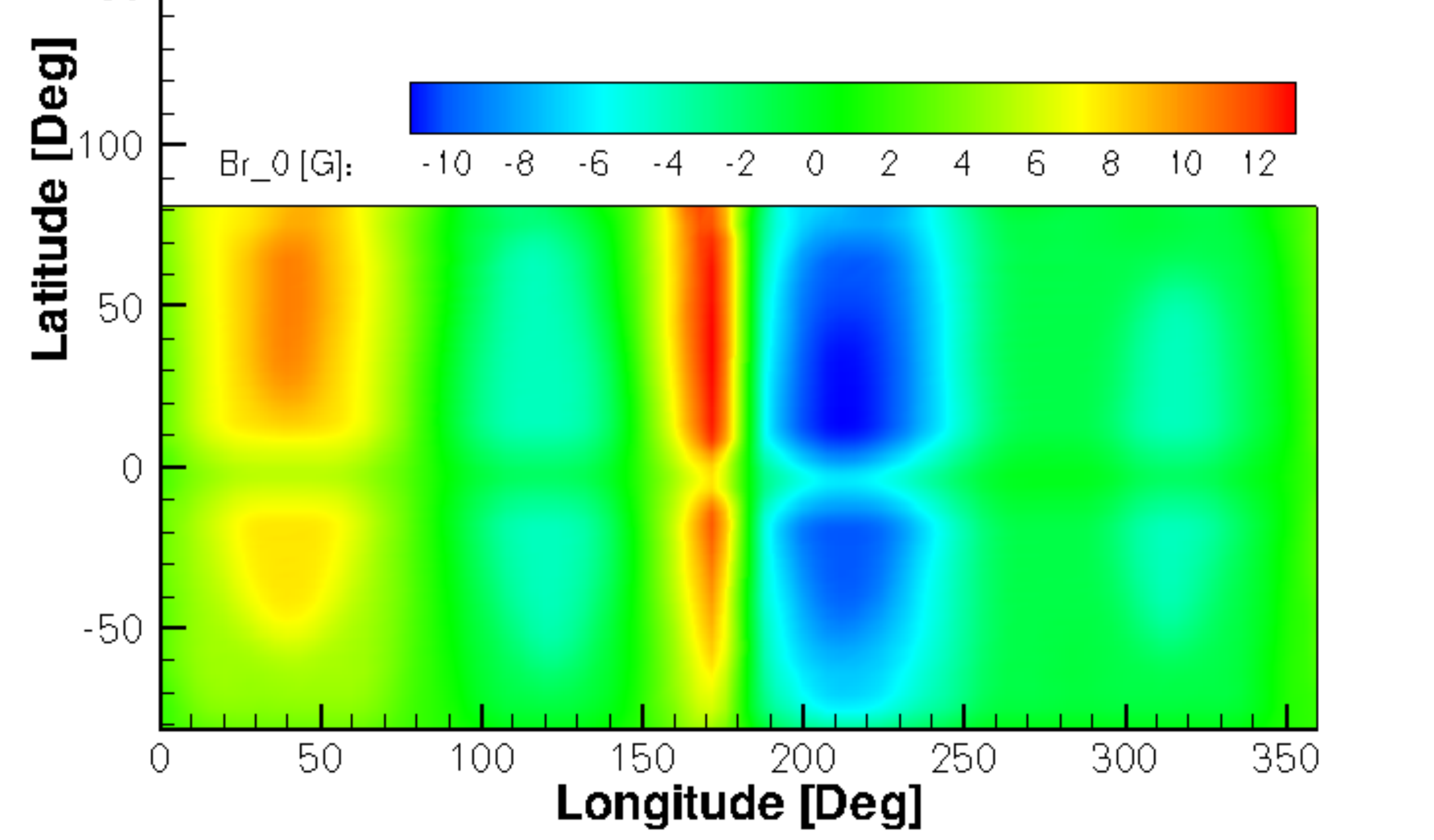} \\
\includegraphics[trim = 0.3in 0.3in
  0.5in 3.5in,clip, width = 0.4\textwidth]{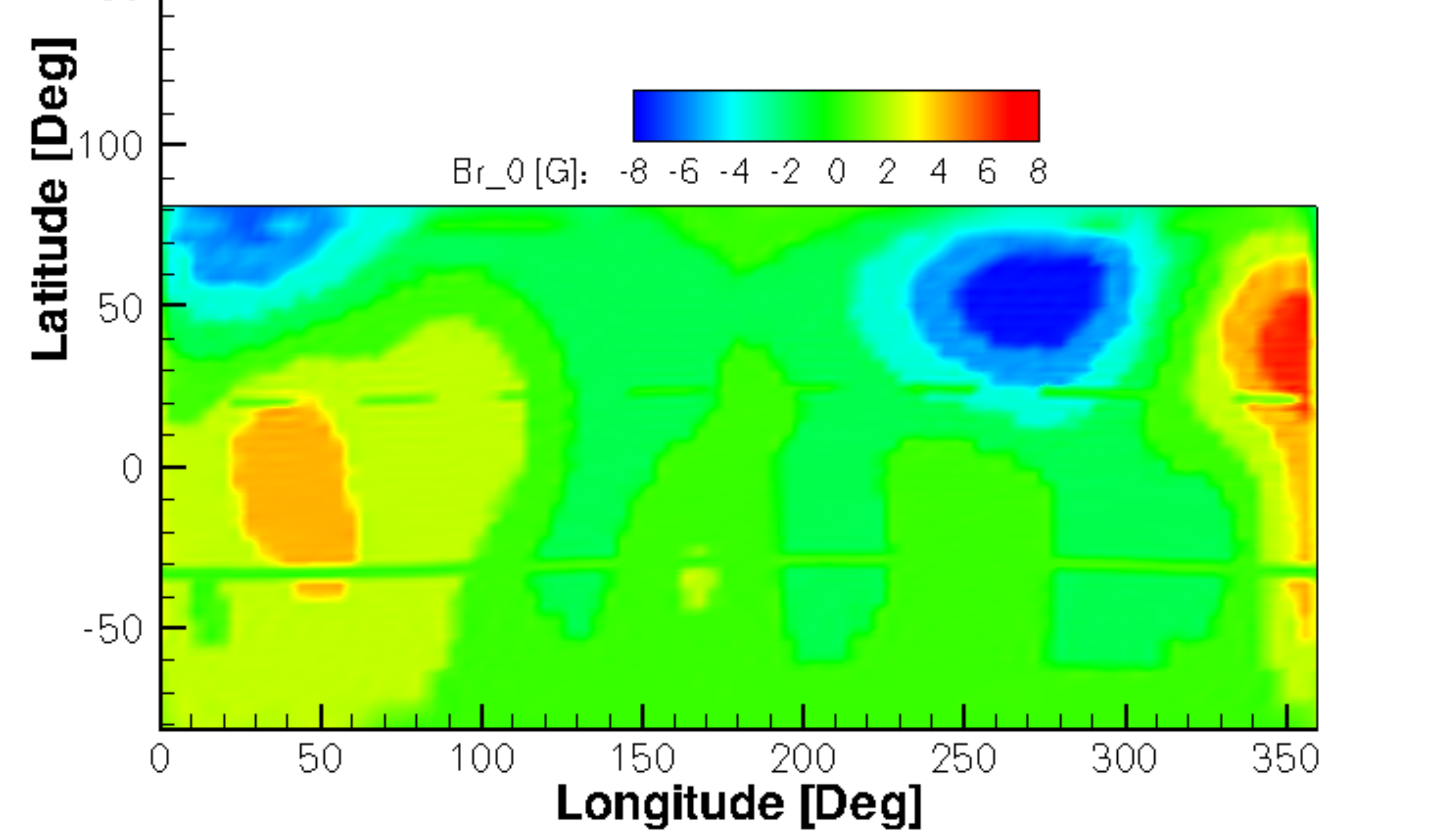} 
\includegraphics[trim = 0.3in 0.3in
  0.5in 3.5in,clip, width = 0.4\textwidth]{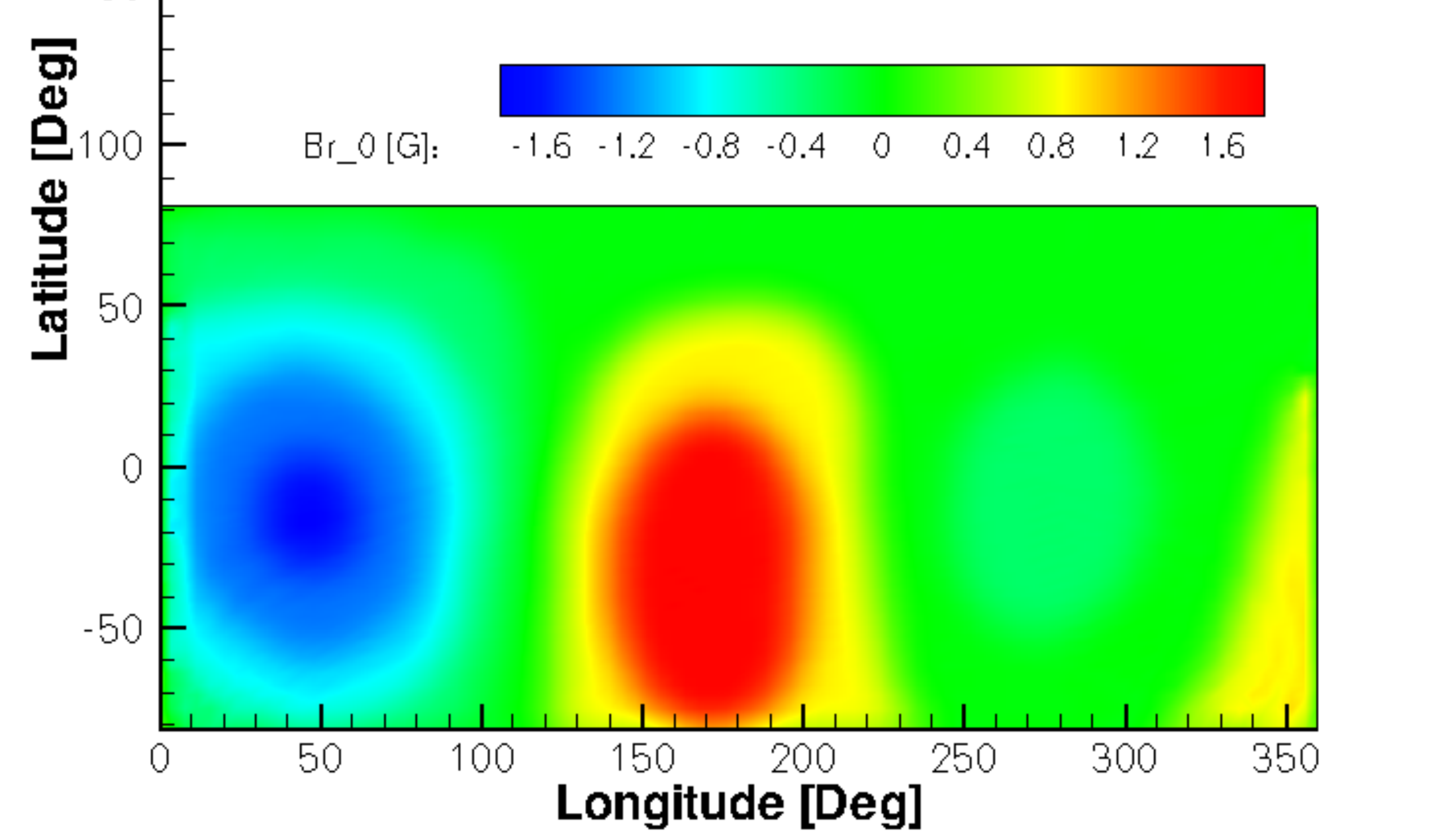} \\
\caption{Magnetograms of the solar analogues in order of ascending age
  from top left to bottom right: HD 29615, HD 35296, AB Dor, HD
  206860, HD 73350, HD 73256, Tau Boo, HD 76151}
\label{fig:magnetograms}
\end{figure*}

\begin{figure*}
\center
\includegraphics[trim = 0.3in 0.3in
  0.5in 3.5in,clip, width = 0.4\textwidth]{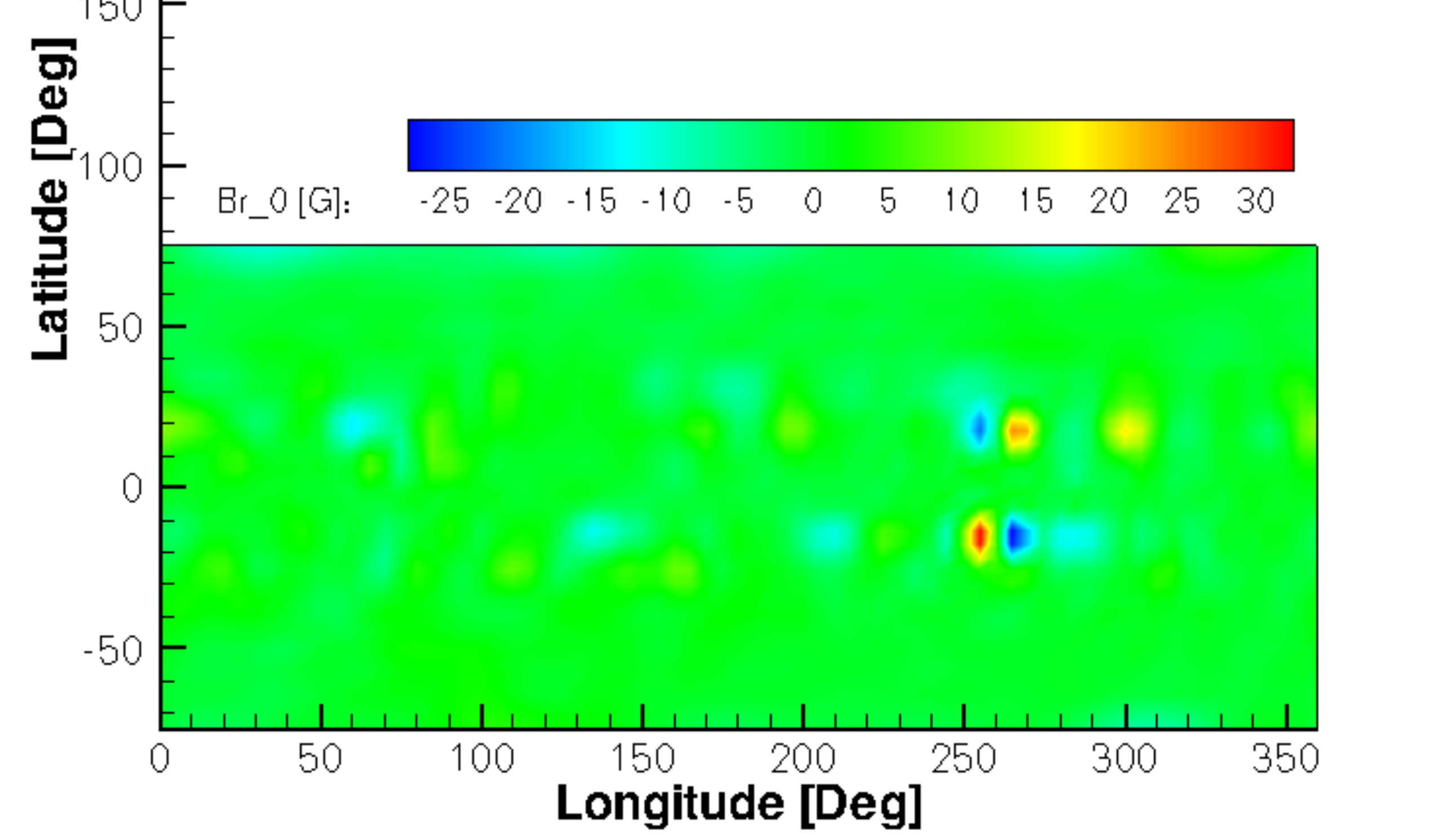} 
\includegraphics[trim = 0.3in 0.3in
  0.5in 3.5in,clip, width = 0.4\textwidth]{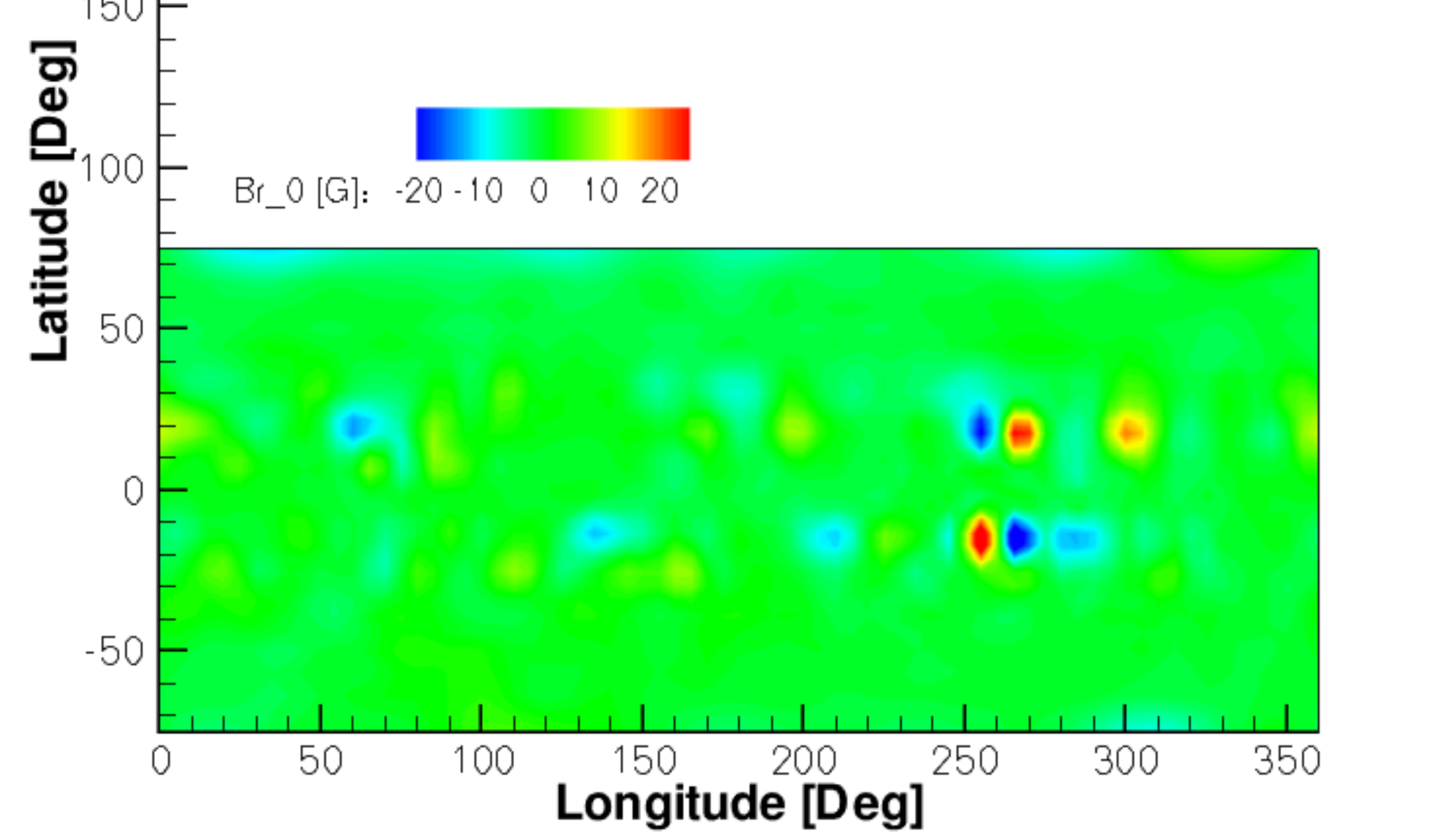}\\ 
\caption{High and low resolution solar magnetograms. High resolution
  was obtained from http://hmi.stanford.edu/data/synoptic.html and low
resolution from http://wso.stanford.edu. Note the different scales.}
\label{fig:solmagnetograms}
\end{figure*}


\section{MHD NUMERICAL SIMULATION}
\label{sec:Methods}

The numerical solutions discussed in this paper are obtained using the
generic {\it BATS-R-US} code \citep{Powell:99}, which is a part of the
Space Weather Modelling Framework \citep[SWMF][]{Toth.etal:12}. The SWMF is
a collection of physics-based models which work in different regimes
of solar and space physics. The results presented
in this paper make use of the solar corona (SC) component \citep[see][]{Vanderholst.etal:14}. To obtain
these results, we use the code to solve a set of 3D MHD
equations for loss of angular momentum and mass. Other parameters such
as maximum and minimum radial wind velocities and their associated
densities are also extracted. The SC module uses maps of the radial magnetic field of the
source to specify the field boundary
conditions \citep[see][for details]{Cohen.etal:08,Alvarado-Gomez.etal:16} at the surface of the star. These maps are obtained using
ZDI, which allows large scale magnetic
topology to be recovered
\citep[e.g.][]{Donati.Brown:97,Hussain.etal:09}. This method has been
thoroughly tested and found to recover the general field distribution of
solar-like stars \citep{Donati.etal:08,Petit.etal:08,Alvarado-Gomez.etal:15,Hussain.etal:16}, thus making it a reliable provider of magnetic field
maps for the SWMF SC module. We also use a high resolution solar
magnetogram obtained using MDI \citep[see][]{Scherrer.etal:95} from
publicly available Stanford data\footnote{http://hmi.stanford.edu/data/synoptic.html}, and a
low resolution magnetogram obtained from the Wilcox Solar Observatory (WSO)\footnote{http://wso.stanford.edu/}. Both solar magnetograms were
taken during the same solar maximum on April 19th 2000.

Once the SC module is provided with the ZDI magnetogram, a field potential
extrapolation above the stellar surface is conducted and is used as
the initial conditions for the simulation. The solar corona is taken to start at the base
of the chromosphere and extends generally up to around~$40R_*$,
though this distance can be larger for stars with strong magnetic
fields. The SC module
also requires input of stellar parameters radius $R_*$, mass $M_*$ and rotation
period $P_{rot}$, as well as the chromospheric base density $n_0$ and
temperature $T_0$. The stellar parameters are obtained from the same
papers as the magnetograms, or failing that are taken to be solar
values. The coronal base density and temperature are taken to be solar values:
$n_0 = 2.0\times10^{8}cm^{-3}$ and $T_0 = 1.5\times10^{6}K$. 
This assumption allows for easier
comparison between the solar analogues and the two solar magnetograms and should be 
a reasonable approximation since all the stars in this sample are solar-like and well below X-ray
saturation. 

We use a spherical three-dimensional grid with a logarithmic scale in the $\hat{r}$
direction. With the magnetic field initial conditions provided by the
magnetogram, the program evolves the coronal heating and stellar winds
self-consistently using Alfv{\'e}n wave dissipation through a turbulent
energy cascade. Processes such as electron heat conduction and
radiative cooling are also taken into account \citep[see][for
details]{Oran.etal:13,Sokolov.etal:13,Vanderholst.etal:14}. We conduct three-dimensional simulations 
for eight solar-like stars and two solar cases, the magnetograms for which can be seen in
Figures \ref{fig:magnetograms} and \ref{fig:solmagnetograms}. The parameters and sources of the
magnetograms can be found in Table ~\ref{tab:magnetograms}.

The simulation provides a three-dimensional solution of the stellar
corona, from which we extract the wind density $\rho$, and the maximum
and minimum radial
velocities, $U_R^{max}$ and $U_R^{min}$. This is
defined as the region of the outer solar corona where the winds speeds are
greater than the Alfv{\'e}n speed,  $v_A=B/\sqrt{4\pi\rho}$,
where $B$ is defined as $B=\sqrt{B_x^2+B_y^2+B_z^2}$ and $\rho$ is the
mass density for a pure hydrogen wind with the electron contribution neglected. The super-Alfv{\'e}nic region is conveniently visualised as the volume
exterior to the Alfv{\'e}n Surface, itself defined as the collection of points in space 
at which the ratio of wind speed to local Alfv{\'e}n speed is unity:
$M_A=U/v_A=1$. We find the total angular momentum and mass
loss rates by integrating the loss rates over the entire
surface. Equations 1 and 2 give the rates for mass
and angular momentum loss respectively:
\begin{equation}
 \frac{dM}{dt} = \rho \, (\mathbf{u} \cdot \mathbf{dA})
 \label{eqn:massloss}
\end{equation}

\begin{equation}
\frac{dJ}{dt} = \Omega \, \rho \, R^2 \sin^2{\theta} \,(\mathbf{u} \cdot \mathbf{dA}),
\label{eqn:angmomloss}
\end{equation}
where $\mathbf{dA}$
is the surface element on the Alfv{\'e}n Surface and $dJ/dt$ is the component of the angular momentum change in the
direction of the rotation axis. We take $dJ/dt$ to be the only
angular momentum component contributing to a
change in the magnitude of $J$. From $dJ/dt$ we may also find the spin down evolution using:
\begin{equation}
\frac{dJ}{dt} = I\frac{d\Omega}{dt},
\label{eqn:spindown}
\end{equation}
where $I$ is the moment of inertia and $\Omega$ is the angular frequency. Integrating a power law fit to $dJ/dt$ will allow a relation for $\Omega$ as a function of time to be obtained and compared to the Skumanich Law. We take the moment of inertia of solar like stars to be $I=0.076MR^{2}$~\citep{Claret.Gimenez:89} and constant in time.

We also find the range of wind speed and density at a distance of 1~AU from the
star, in order to study the effect of ram pressure on a magnetosphere with a magnetic field similar to that of a young Earth's or a similarly placed planet around
a solar analogue. A maximum and minimum wind speed and their associated
densities are found within the stellar corona solution, in the super-Alfv{\'e}nic
region. As the wind speeds asymptotically approach a maximum at the
Alfv{\'e}n Surface, the maximum and minimum found in the stellar corona solution
in the super-Alfv{\'e}nic region can be used directly for the 1~AU distance. The corresponding densities
are extrapolated following a $1/{R^2}$ law. 
In order to find an approximate range of values for the ratio of the size of a test planet's
magnetosphere to the planet's radius, we equate wind ram pressure to the 
magnetic pressure of the planetary field and rearrange to get the usual relation for the magnetosphere standoff distance when the wind magnetic pressure can be neglected:

\begin{equation}
\frac{R_{Mp}}{R_p} = \left(\frac{2 B_p^2}{ \mu_0 \,\rho \,U_R^2}\right)^{\frac{1}{6}}
\end{equation}
where $\mu_0$ is the permeability of free space and $B_p$ is the
equatorial magnetic field strength at the planet's surface. We also extrapolate ram pressure up to 100~Au for a range of stellar ages, in order to give a clear visual of the decline of ram pressure over time.\\

\begin{table*}
\begin{center}
\caption{Parameters of stellar sample and magnetogram sources.}
\label{tab:magnetograms}
\begin{tabular}{c c c c c c }
\tableline
 & Spectral & Age & $P_{rot}$ & $Log(L_X)$\tablenotemark{b} & ZDI \\
Star & Type & (Gyr) & (days) & ($erg/s/cm^2$) & Ref. \\
\tableline
HD 29615 & G3V & $0.03_{-0.01}^{+0.01}$ & $2.34_{-0.05}^{+0.02}$ &  & 1 \\ 
HD 35296 & F8V & $0.04_{-0.01}^{+0.01}$  & $3.48_{-0.01}^{0.01}$ & 29.33 & 1 \\ 
AB Doradus & K0V & $0.075_{-0.025}^{+0.025}$ & $0.5_{-0.1}^{+0.1}$ & 30.33 & 2 \\ 
HD 206860 & G0V & $0.25_{-0.05}^{+0.05}$ & $4.6_{-0.1}^{+0.1}$ & 29.19& 3 \\ 
HD 73350 & G5V & $0.80_{-0.30}^{+0.30}$ & $12.3_{-0.1}^{+0.1}$ & 28.72 & 4 \\ 
HD 73256 & G8 & $0.83_{-0.03}^{+0.03}$ & $14.0_{-0.1}^{+0.1}$ &  & 5 \\ 
$\tau$ Bo{\"o}tis & F7V  & $2.4_{-1.1}^{+0.7}$ & $3.0_{-0.1}^{+0.1}$ & 28.85 & 6 \\ 
HD 76151 & G3V & $3.6_{-2.3}^{+1.8}$ & $20.5_{-0.3}^{+0.3}$ & 28.40 & 4 \\ 
Sun (high res)\tablenotemark{a} & G2V & $4.57_{-0.01}^{+0.01}$ & $25.5_{-0.01}^{+0.01}$ & 27.3\tablenotemark{c} & 7 \\
Sun (low res)\tablenotemark{a} & G2V & $4.57_{-0.01}^{+0.01}$ & $25.5_{-0.01}^{+0.01}$ & 27.3\tablenotemark{c} & 8\\
\tableline
\end{tabular}
\end{center}
\tablerefs{(1) \cite{Waite.etal:15}; (2) \cite{Hussain.etal:07}; (3) \cite{BoroSakia.etal:15}; (4) \cite{Petit.etal:08}; (5) \cite{Fares.etal:13}; (6) \cite{Fares.etal:09}; (7) http://hmi.stanford.edu/data/synoptic.html; (8) http://wso.stanford.edu/ }
\tablenotetext{a}{Corresponding to the solar maximum of 2000 April 19.}
\tablenotetext{b}{X-ray flux from http://www.hs.uni-hamburg.de/DE/For/Gal/Xgroup/nexxus/index.html}
\tablenotetext{c}{From \citet{Gudel.etal:96}}
\end{table*}

\begin{figure*}
\center
\includegraphics[trim=0.1in 0.1in
  0.1in 0.1in,clip,width=0.8\textwidth]{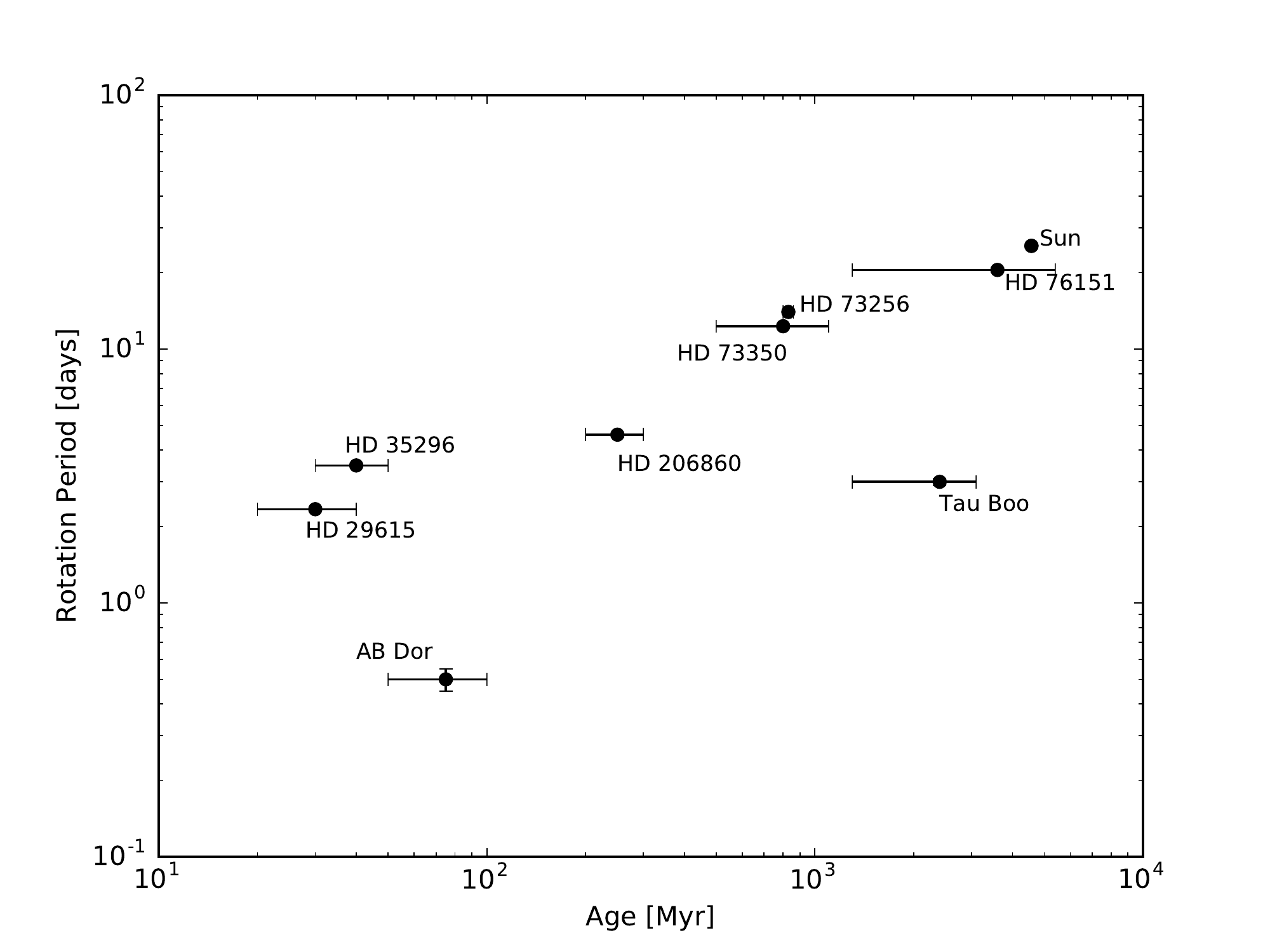}
\caption{Plot of rotation period as a function of age for the stars in our sample. Note that the error bars for rotation period are too small to be seen.}
\label{fig:agerot}
\end{figure*}


\section{STELLAR SAMPLE}
\label{sec:sample}

We use a sample of 8 solar-like stars and 2 solar magnetograms of different resolutions to conduct our investigation. The stellar parameters and magnetogram sources are described in brief in table \ref{tab:magnetograms}, and individually in more detail in the following subsections. While our sample contains a range of spectral types from F7 to K0, our assumption is that we can use the model honed to match the observed solar wind for any star with a hot corona for which the stellar wind is accelerated in a similar manner to the solar wind \citep{Cohen.etal:10a,Cohen.etal:12}. This is in contrast to, for example, line driven winds in giant stars (see e.g. \citet{Airapetian.etal:03,deBeck.etal:10}). Our model assumes coronal heating and wind acceleration via Alfv{\'e}n wave dissipation, while also taking into account the scaling of the magnetic flux with X-ray flux \citep{Pevtsov.etal:03} via the Poynting flux providing the energy to the corona. This implementation accounts, at least in part, for the affect of different convection zone parameters on the coronal heating and wind acceleration. The latter relation is worthy of further study, though falls beyond the scope of the present work.

\subsection{HD 29615}
\label{sec:HD29615}
HD~29615 is a G3V type star of age $30_{-10}^{+10}$ Myr \citep{Zuckerman.Song:04,Waite.etal:15} with a rotation period of $2.34_{-0.05}^{+0.02}$ days \citep{Waite.etal:15}.

\subsection{HD 35296}
\label{sec:HD35296}
HD~35296 is a F8V type star of age $40_{-10}^{+10}$ Myr, with a rotation period of $3.48_{-0.01}^{+0.01}$ days \citep{Waite.etal:15}.

\subsection{AB Dor}
\label{sec:ABDOR}

AB Dor is a K0 dwarf \citep{Torres.etal:06} of age 50 Myr, with a rotation period of $0.5_{-0.1}^{+0.1}$ days \citep{Hussain.etal:07,Plavchan.etal:09}. While it is clear that AB Dor must be a young star, the exact age is subject to some uncertainty. We adopt the value of $75_{-25}^{+25}$ Myr \citep{Zuckerman.Song:04,Plavchan.etal:09} as a fairly conservative estimate and range.
There is some debate as to whether AB Dor is a pre-main sequence star or if it has just arrived on the main sequence \citep{Hussain.etal:07}. While differing in spectral type from the Sun more that the  other stars of the study, AB~Dor still represents the most well-studied rapidly rotating solar-like star, especially in regard to surface magnetic field distribution, activity and winds (see, e.g., \citet{Hussain.etal:07,Donati.etal:09,Cohen.etal:10b}).

\subsection{HD 206860}
\label{sec:HD206860}

HD 206860 (also known as HN Peg) is a G0 type star of age 200 Myr, with a rotation period of $4.6_{-0.1}^{+0.1}$ days \citep{Pizzolato.etal:03,BoroSakia.etal:15}. There has been some matter of debate over the true age of HD 206860. \citet{Luhman.etal:07} give an age of 300 Myr, while \citet{Barnes:07} finds $247_{-42}^{+42}$ Myr using gyrochronology. We decide to take a conservative middle ground and use $250_{-50}^{+50}$ Myr as HD 206860's age.

\subsection{HD 73350}
\label{sec:HD73350}

HD 73350 is a G5 type star with a rotation period of $12.3_{-0.1}^{+0.1}$ days \citep{Petit.etal:08,Vidotto.etal:14a}. HD 73350 was originally thought to be quite an old star only slightly younger than the Sun, with an age of $4.1_{-2.7}^{+2.0}$ Gyr \citep{Valenti.Fischer:05}, but more recent studies have observed a debris disk, yielding a much younger age of $513_{-136}^{+136}$ Myr \citep{Plavchan.etal:09}. \textbf{However, this appears to be an underestimation, as a recent survey of the Hyades cluster ($\sim$800 Myr) has found similar G5 stars with rotation periods of $\sim$10-11 days \citep{Douglas.etal:16}. In order to accurately reflect the uncertainty in this star's age, we take a value of $800_{-300}^{+300}$ Myr. }

\subsection{HD 73256}
\label{sec:HD73256}

HD 73256 is a G8 type star of age $830_{-30}^{+30}$ Myr \citep{Udry.etal:03,Saffe.etal:05,Fares.etal:13} with a rotation period of $14.0_{-0.1}^{+0.1}$ days \citep{Udry.etal:03,Fares.etal:13}. It should be noted that HD 73265 has a hot Jupiter orbiting it with a period of about 2.55 days at a distance of 0.37 Au \citep{Udry.etal:03}. It is possible that the presence of a close-in hot Jupiter may affect the stellar activity properties through influence on rotation, though no evidence of tidal effects have been seen, and the stellar rotation and planet orbital periods are very different. 

\subsection{$\tau$~Bo{\"o}tis}
\label{sec:TauBoo}

$\tau$~Bo{\"o}tis is a F7 type star of age 2.4 Gyr \citep{Saffe.etal:05,Mamajek.Hillenbrand:08}, with a rotation period of $3.0_{-0.1}^{+0.1}$ days \citep{Fares.etal:09}. $\tau$~Bo{\"o}tis' age has also been hotly debated, and is often cited to be anywhere between 1.3-3.1 Gyr \citep[e.g.][]{Fuhrmann.etal:98, Borsa.etal:15}; we chose to use $2.4_{-1.1}^{+0.7}$ Gyr. Furthermore, $\tau$~Bo{\"o}tis hosts a hot Jupiter with an orbital period of 3.3 days \citep{Butler.etal:97,Catala.etal:07,Donati.etal:08, Fares.etal:09}. \citet{Donati.etal:08} even suggest that "the tidal effects induced by
the giant planet can be strong enough to force the thin convective envelope into corotation." Thus, while we include $\tau$~Bo{\"o}tis in our sample as we still believe the wind producing mechanism is solar like, the MHD simulation results for this star should be taken as potentially less reliable than for the rest of the sample.

\subsection{HD 76151}
\label{sec:HD76151}

HD 76151 is a G3 type star of age $3.6_{-2.3}^{+1.8}$ Gyr \citep{Valenti.Fischer:05,Vidotto.etal:14a}, with a rotation period of $20.5_{-0.3}^{+0.3}$ days \citep{Petit.etal:08,Vidotto.etal:14a}.

\subsection{Sun}
\label{sec:Sun}

The Sun is a G2 type star of age $4.57_{-0.01}^{+0.01}$ Gyr, with a rotation period of $25.5_{+0.1}^{-0.1}$ days \citep{Bonanno.etal:02,Pizzolato.etal:03,Connelly.etal:12}. While the underlying magnetic morphology of the Sun is normally a dipole, this is not the case during a solar maximum such as that of April 2000 when our two solar magnetograms were taken.

\begin{table*}
\begin{center}
\caption{Calculated quantities from MHD simulations}
\label{tab:results}
\begin{tabular}{c c c c c c }
\tableline
 & $\left\langle B\right\rangle$\tablenotemark{b} & dM/dt\tablenotemark{c} & dJ/dt\tablenotemark{d} & $\rho (U_R^{max})^{2}$ & $\rho (U_R^{min})^{2}$\\
Star & (G) & ($10^{-14} M_{sol}/yr$) & (erg) & ($10^{-8} dyn~cm^{-2}$)\tablenotemark{e} & 
($10^{-8} dyn~cm^{-2}$)\tablenotemark{e} \\
\tableline
HD 29615 & 6.85E1 & 9.50 & 4.69E31 & 23.6 & 42.9 \\ 
HD 35296 &7.16 &3.18 & 3.00E31 &4.42 &5.77  \\ 
AB Doradus & 1.19E2& 12.5& 2.15E33& 43.5 &48.1  \\ 
HD 206860 &6.64& 2.27& 2.57E31 &7.07 &10.9  \\ 
HD 73350 & 2.45& 1.10& 2.53E30& 1.28 &2.82  \\ 
HD 73256 &2.96& 0.745& 1.26E30 &7.16 &5.72  \\ 
$\tau$ Bo{\"o}tis &1.26 &0.933 & 1.19E31 &2.07 &1.45  \\ 
HD 76151&0.42& 0.210& 9.32E28 &0.331 &0.253  \\ 
Sun (high res)\tablenotemark{a} &1.46& 0.539& 1.14E29 &1.97 &0.995  \\
Sun (low res)\tablenotemark{a} &1.46& 0.563& 1.34E29 &1.62 &0.758 \\
\tableline
\end{tabular}
\end{center}
\tablenotetext{a}{Corresponding to the solar maximum of 2000 April 19.}
\tablenotetext{b}{Mean surface magnetic field computed from the magnetogram.}
\tablenotetext{c}{Mass loss rate}
\tablenotetext{d}{Angular momentum loss rate}
\tablenotetext{e}{Wind ram pressure}
\end{table*}

\begin{figure*}
\center
\includegraphics[trim = 0.1in 0.1in
  0.1in 0.1in,clip, width = 0.7\textwidth]{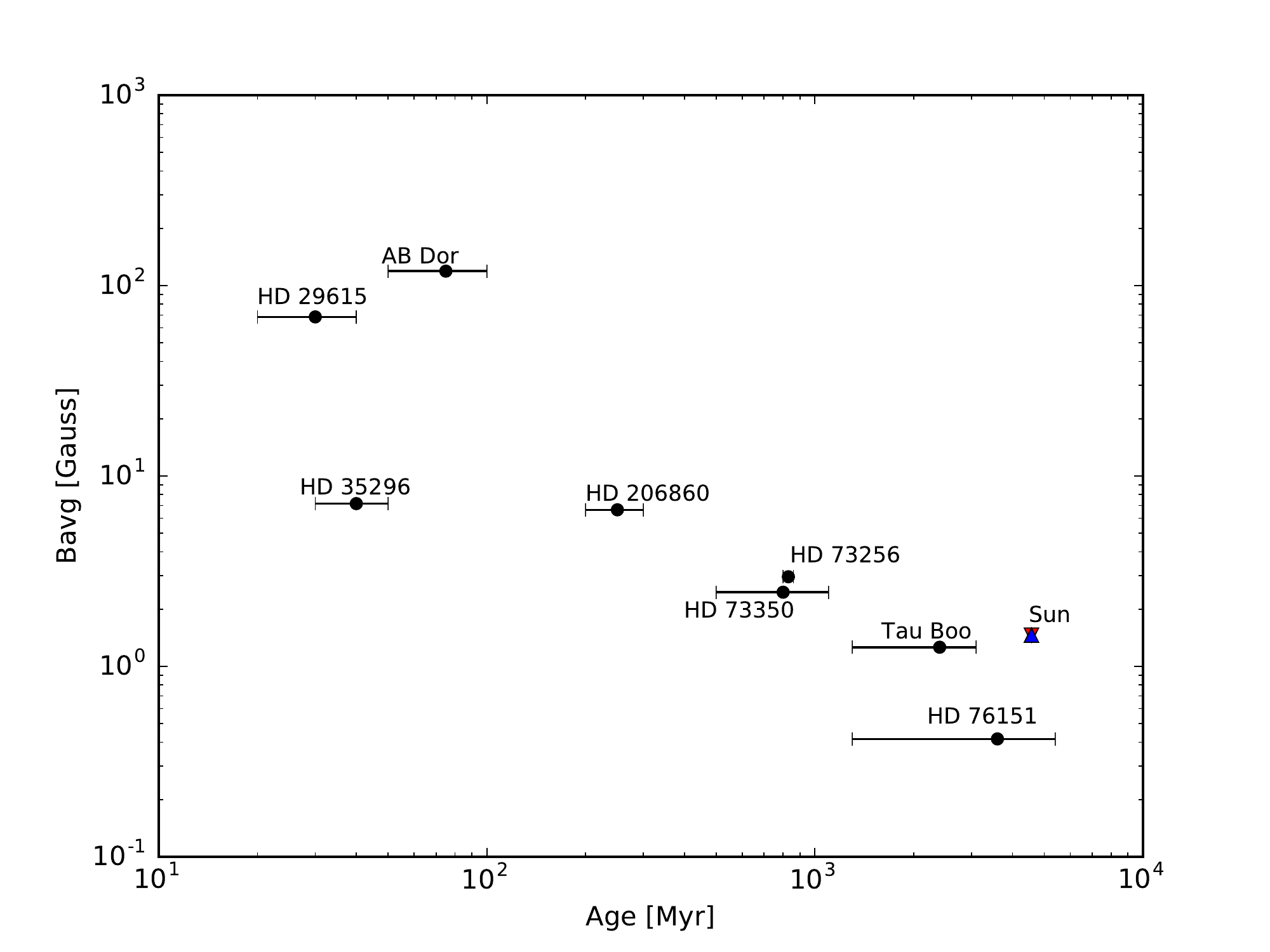} 
\includegraphics[trim = .1in .1in 
0.1in 0.1in,clip,width =0.7\textwidth]{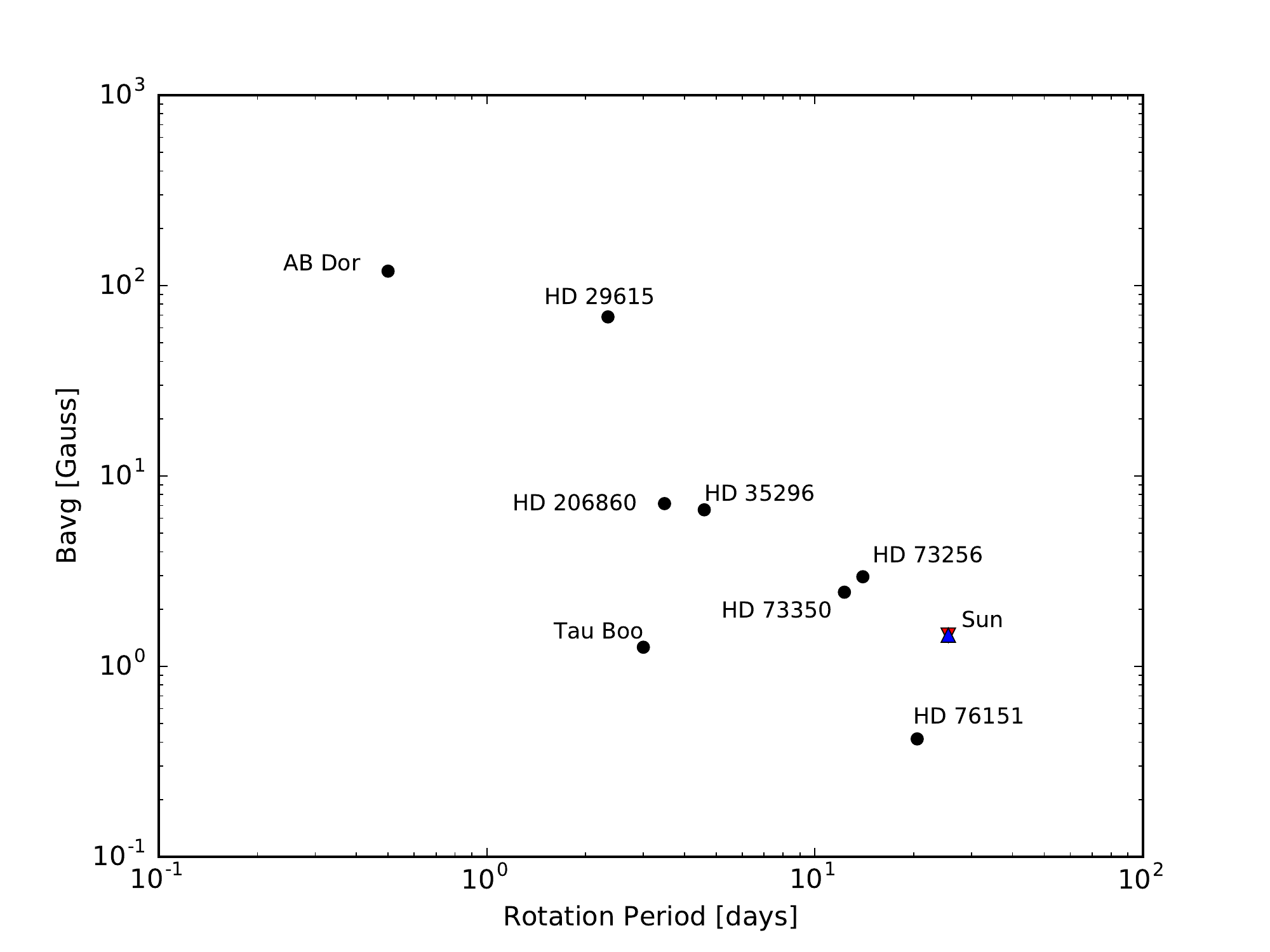}\\
\caption{Plots of average surface magnetic field strength against stellar age and rotation period.
The upward triangle is the high resolution solar case and the
downward triangle the low resolution solar case. Note that these overlap.}
\label{fig:Bavg}
\end{figure*}

\begin{figure*}
\center
\includegraphics[trim = 0.1in 0.1in
  0.1in 0.1in,clip, width = 0.7\textwidth]{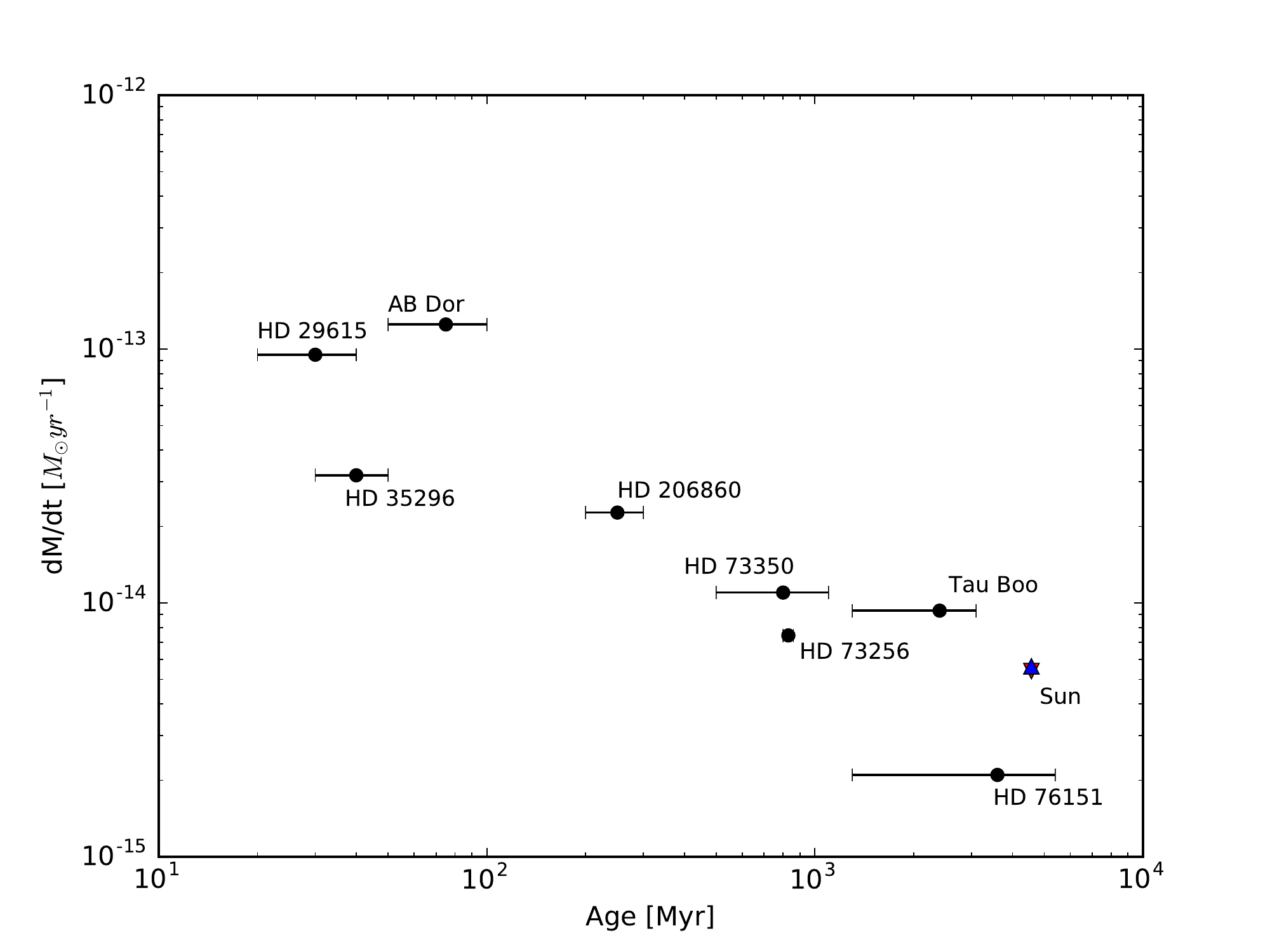} 
\includegraphics[trim = .1in .1in 
0.1in 0.1in,clip,width =0.7\textwidth]{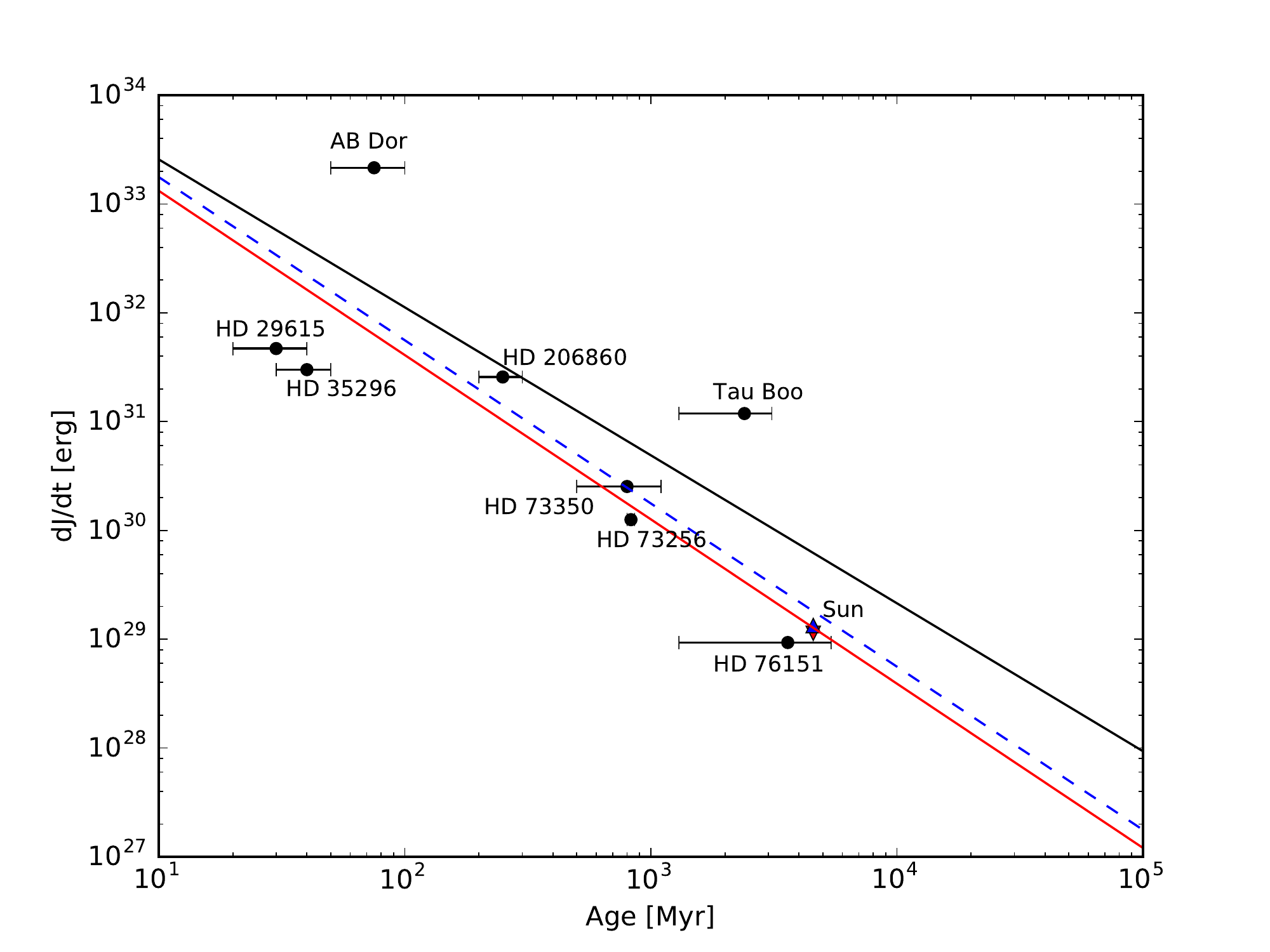}\\
\caption{Plots of mass and angular momentum loss rates against stellar age.
The upward triangle is the high resolution solar case and the
downward triangle the low resolution solar case. In the bottom plot, the blue dashed line is the Skumanich Law, while the black line is our power law fit with $\Omega\sim t^{-0.36}$. The red line is our power law fit excluding $\tau$~Bo{\"o}tis with $\Omega\sim t^{-0.51}$.}
\label{fig:age}
\end{figure*}

\section{RESULTS}
\label{sec:Results}

We present the results for the solar analogues in section \ref{sec:analogues},
including a look at their age-rotation trend (Figure \ref{fig:agerot}) as well as their average surface magnetic fields with respect to age and rotation period (Figure \ref{fig:Bavg}). We also look at the mass and
angular momentum loss rates (Figures \ref{fig:age} and \ref{fig:rot}) and calculate a range of possible
magnetosphere sizes for a sample planet with a 0.3G equatorial field (Figure \ref{fig:Rmp}). Ram pressures are also extrapolated up to 100~Au for locii of different ages (Figure~\ref{fig:Rampress}). In
Section \ref{sec:solresults} we look at the results from the two solar magnetograms,
and compare these to the ones obtained from the solar
analogues. Three-dimensional plots of the stellar coronae complete
with Alfv{\'e}n Surface, wind velocity slice and magnetic field lines can
be found in Figures ~\ref{fig:3dvisa} and ~\ref{fig:3dvisb}, to give a clearer visualisation of simulation results. The numerical results are summarised concisely in Table \ref{tab:results}.

\subsection{Solar Analogues}
\label{sec:analogues}

 
The rotation periods of the stars of our sample are illustrated as a function of their ages in Figure~\ref{fig:agerot}. The trend shown follows the expected age-rotation relation, where the period increases as the
stars get older \citep[e.g.][]{Gallet.Bouvier:13}.  As the sample demonstrates, our study covers a significant range of rotation
periods for both younger and older stars, with AB Doradus rotating with a period of only 
0.5 days at a young age of 75~Myr, and HD 76151 rotating with a period of 20.5 days at an age of 3.6~Gyr. We also note that $\tau$~Bo{\"o}tis has a period of only 3 days \citep{Donati.etal:08,Fares.etal:09} although it is thought to be fairly old at 2.4~Gyr \citep{Saffe.etal:05}. If we look at the four
stars older than 500~Myr, it is apparent that $\tau$~Bo{\"o}tis is indeed somewhat
of an outlier for its age. However, it should be noted that this star has a hot Jupiter companion which could affect the magnetic cycle, activity and angular momentum loss \citep[see][for details]{Butler.etal:97,Catala.etal:07,Donati.etal:08,Fares.etal:09,Cohen.etal:10b}.
In Figure \ref{fig:Bavg}, we observe a decrease of magnetic field strength with age and rotation period. Since stellar magnetic dynamos are thought to be intrinsically linked to rotation \citep[e.g.][]{Skumanich:72,Pallavicini.etal:81,Wright.Drake:11}, this an expected result \citep[see also, e.g.,][]{Vidotto.etal:14a}. 

\begin{figure*}
\center
\includegraphics[trim = 0.1in .1in
  0.1in 0.1in,clip, width = 0.7\textwidth]{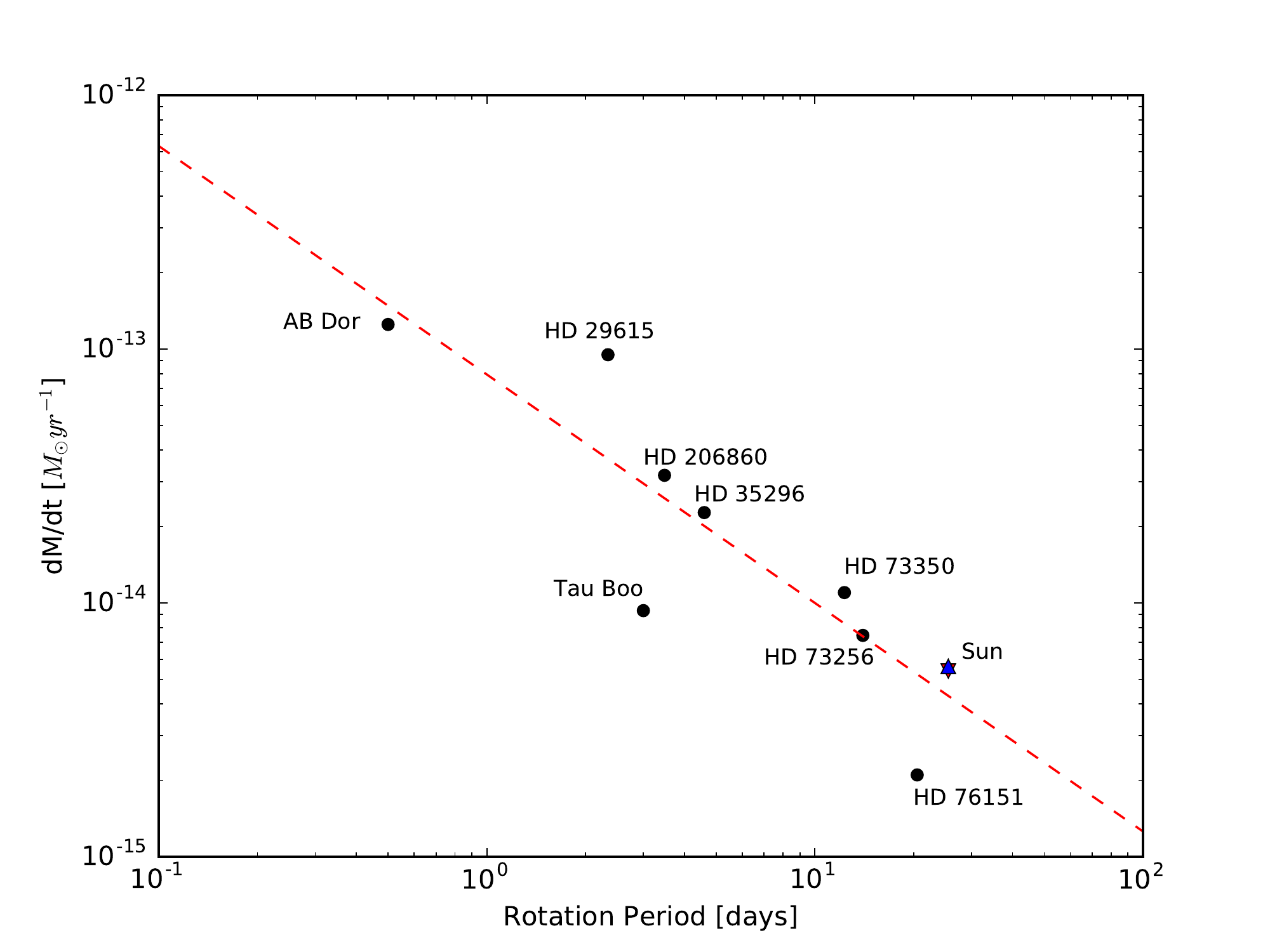} 
\includegraphics[trim = .1in .1in 
0.1in 0.1in,clip,width = 0.7\textwidth]{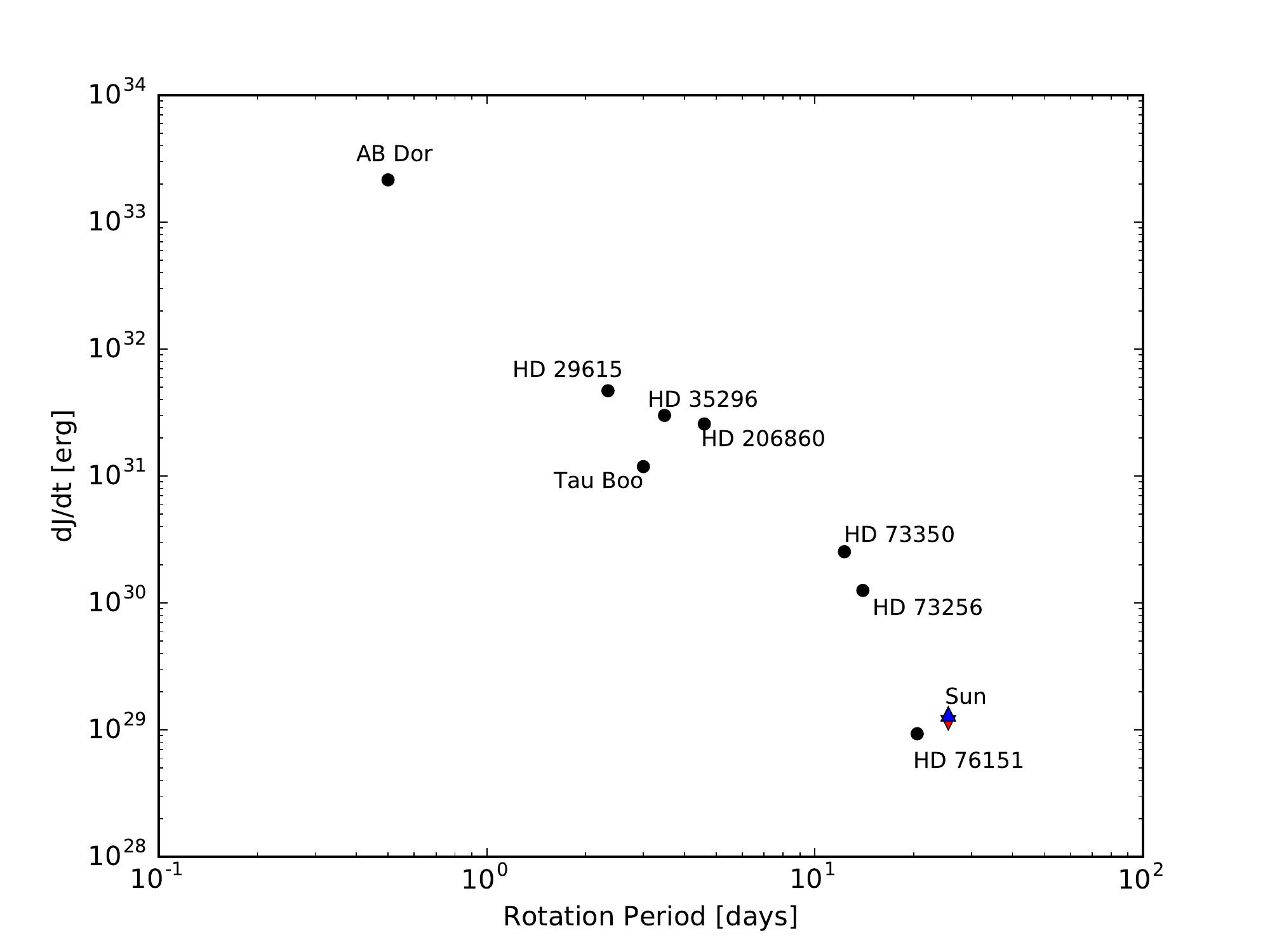}\\
\caption{Plots of mass and angular momentum loss rates against 
  rotation period. The upward triangle is the high resolution solar case and the
downward triangle the low resolution solar case (note that these symbols overlap on the plot). The red line in the top plot is our fit of $dM/dt \propto P^{-0.90}$. }
\label{fig:rot}
\end{figure*}

Mass and angular momentum loss rates as a function of both stellar age and rotation period are illustrated in Figures~\ref{fig:age} and \ref{fig:rot}, respectively.  Both quantities exhibit descending trends with age and rotation period, as expected.  Again, $\tau$~Bo{\"o}tis lies above the trends for its age, by almost an order of magnitude for mass loss and two orders of magnitude for angular momentum loss, compared to the 3.6 Gyr old star HD~76151. 

The bottom plot in Figure~\ref{fig:age} illustrating the angular momentum loss rate also shows the Skumanich Law describing the decay of angular velocity with time, $\Omega\sim t^{-0.5}$, in blue and a power law fit to the data.  The latter corresponds to $\Omega\sim t^{-0.36}$ in black. The older stars conform quite closely to the Skumanich relation, with the exception of $\tau$~Bo{\"o}tis. When removing $\tau$~Bo{\"o}tis from the fit, we find a relation of $\Omega \sim t^{-0.51}$ (shown in red in the bottom panel).

Figure~\ref{fig:rot} shows the stars organised by rotation period as opposed to age. We see that $\tau$~Bo{\"o}tis fits quite well with younger stars of comparable period for angular momentum loss rate, but yields a much lower mass loss rate. We discuss this star as an outlier in our sample in Section~\ref{sec:Taudiscuss}. 

\begin{figure*}
\center
\includegraphics[trim = 0.1in .1in
  0.1in 0.1in,clip, width = 0.7\textwidth]{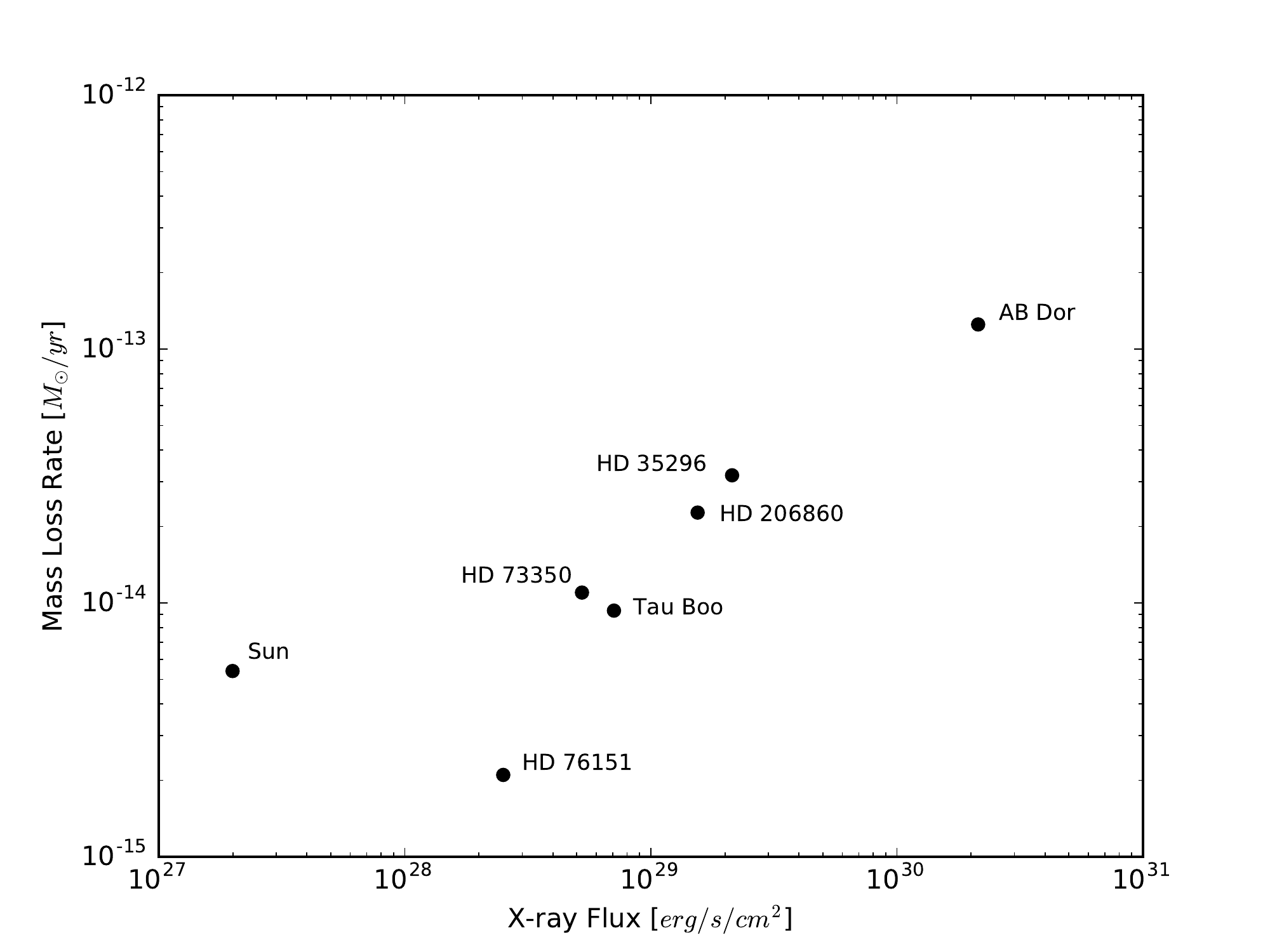} 
\includegraphics[trim = .1in .1in 
0.1in 0.1in,clip,width = 0.7\textwidth]{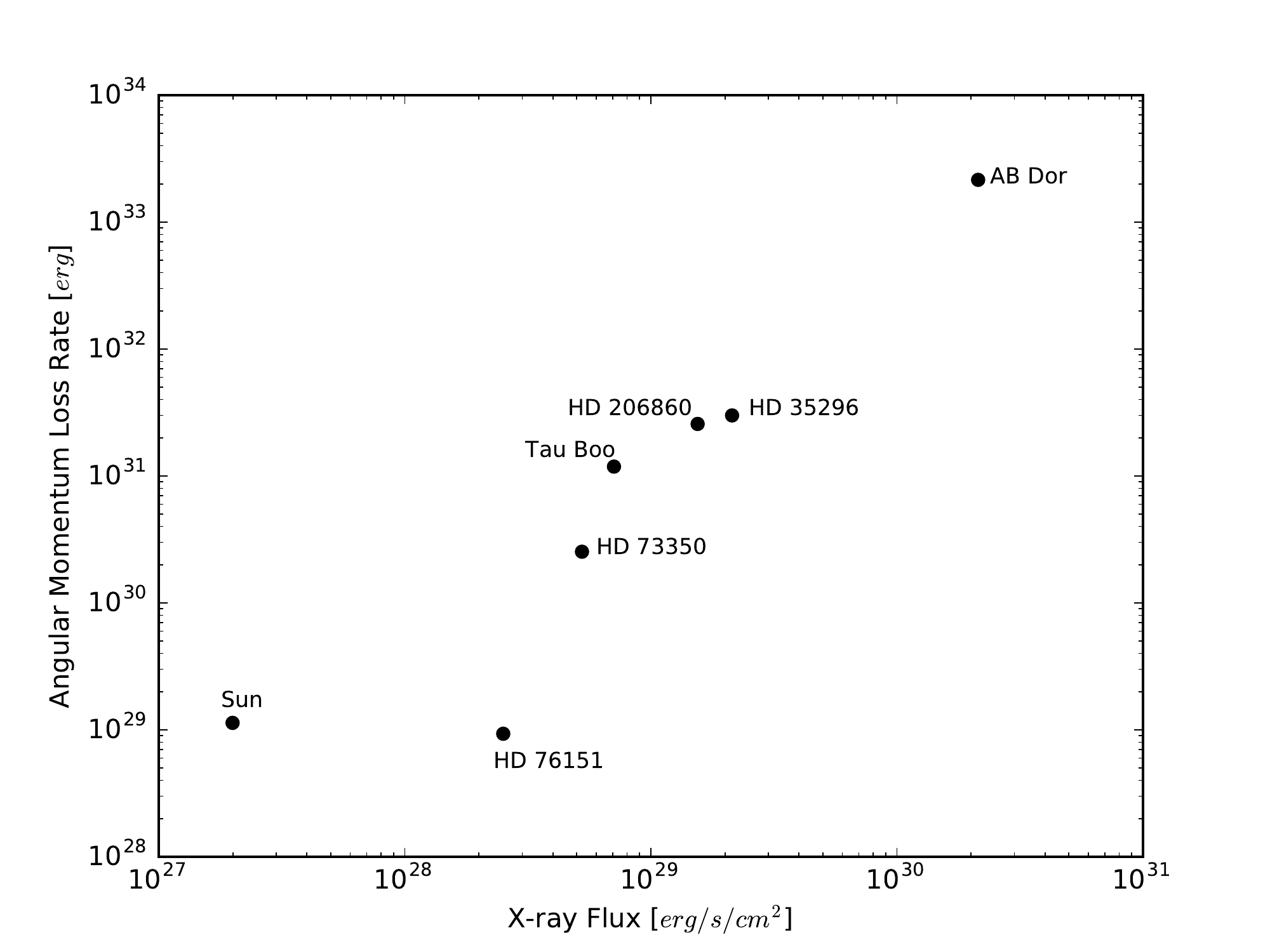}\\
\caption{Plots of mass and angular momentum loss rates against 
 coronal X-ray flux\footnote{X-ray fluxes from \citet{Gudel.etal:96} and http://www.hs.uni-hamburg.de/DE/For/Gal/Xgroup/nexxus/index.html}.}
\label{fig:Lx}
\end{figure*}

It is also known that the coronal X-ray emissions of stars correlate very well with stellar age and rotation, notably for solar type stars (see e.g. \citet{Gudel.etal:96,Ribas.etal:05,Guinan.Engle:09}). X-ray emission is conspicuously greater for younger stars, as are mass and angular momentum loss rates. We thus find a consistent result when plotting these loss rates against X-ray flux in Figure \ref{fig:Lx}, where we see a clear correlation between the quantities.

\begin{figure*}
\center
\includegraphics[trim=0.1in 0.1in
  0.1in 0.1in,clip,width=0.8\textwidth]{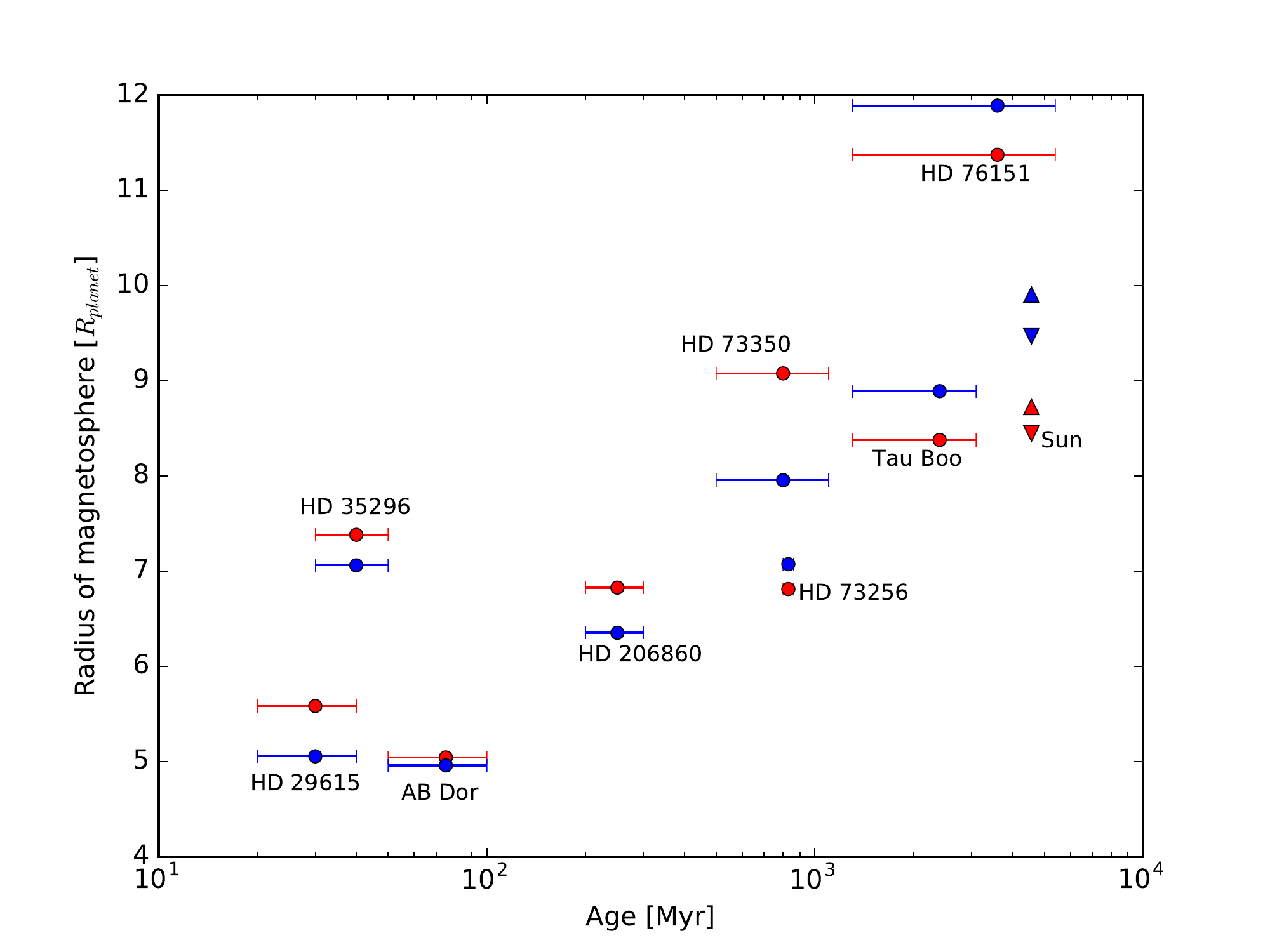}
\caption{Plot of possible magnetospheric standoff distances for an Earth-like planet situated at 1~AU with a 0.3G
  equatorial magnetic field vs. stellar age. The red points represent
  the radius of the magnetosphere for the fastest stellar wind, and
  the blue points for the slowest stellar wind. The upwards triangles
  are the high resolution solar case and the downwards triangle the
  low resolution solar case.}
\label{fig:Rmp}
\end{figure*}

Finally, Figure~\ref{fig:Rmp} illustrates the magnetospheric stand-off height of a 1~AU planet with an equatorial dipolar field strength of 0.3~G as a function of stellar age for fast and slow wind conditions.  
The
general trend is that the magnetospheric radii 
grow larger as the stars grow older, reflecting the commensurate decline in wind intensity. In the case of the 
for younger stars, the fast wind conditions tend to result in a larger
magnetosphere, whereas the opposite is true for our sample of older
stars, where the slow wind conditions yields larger magnetospheres. 

Three-dimensional visualisations of the solar analogue wind conditions are shown in
Figures~\ref{fig:3dvisa} and~\ref{fig:3dvisb}.  In general, 
the younger stars have stronger and somewhat more complex magnetic fields and faster winds than the older stars. The four youngest
stars, HD~29615, HD~35296, AB Doradus and HD~206860, all have maximum wind speeds that reach well 
over 1000~km $s^{-1}$, whilst the older stars HD~73350,
HD~73256, $\tau$~Bo{\"o}tis and HD~76151 barely reach such
velocities. The complexity of the magnetic field is also more
important for the younger stars, which often have twisting and bending
field lines, related to fast rotations as has been shown in other studies \citep[e.g.][]{Cohen.Drake:14}. This transitions to calmer and straighter field lines for
older stars, with HD~206860 being a good example of the transitory
stage. However, it should be noted that the magnetic field of $\tau$~Bo{\"o}tis 
is also quite ordered, even with a relatively rapid
rotation period of 3 days. Owing to their more rapid rotation, younger stars have
stronger magnetic fields and this trend is followed by the stars in
our sample. In particular, the stars HD~29615 and AB~Doradus have very
strong fields, which are responsible for the very large sizes of their
Alfv{\'e}n Surfaces. The domain of the simulations had to be
extended to 60 and 80 stellar radii, respectively, in order to fully
encompass their Alfv{\'e}n surfaces and accurately calculate mass and angular
momentum loss rates.

\subsection{Solar Magnetograms}
\label{sec:solresults}
The mass and angular momentum loss rates of the solar magnetograms 
are included in Figures ~\ref{fig:age} and ~\ref{fig:rot}.
Immediately, it is clear that the loss rates do not depend strongly on magnetogram resolution. 

The mass loss rates are found to be
$5.4\times 10^{-15}$ and $6.0\times 10^{-15} M_\odot$~yr$^{-1}$ for the high and low resolution magnetograms,
respectively. These values are slightly lower than the observed value of about $2\times 10^{-14} M_\odot$~yr$^{-1}$. The explanation for this is that solar wind MHD models typically need a factor of 2--3 applied to the input magnetograms in order to obtain the right magnitude of the magnetic flux at 1~AU \citep{Cohen.etal:08,Riley.etal:12,Linker.etal:12}.  We do not wish to apply arbitrary scaling factors to the stellar magnetograms here as the aim of our study is to examine the trends in wind conditions using only the magnetic field data as an input. We therefore choose to use a scaling factor of 1 for all maps, including the solar ones, for consistency, and as such do not expect to recover solar mass loss rates consistent with observations. Due to the lower magnetic field in the solar map, the Alvf\'en surface is smaller than the one we would have with a scaling factor of 2--3 and the mass loss rate is thus also reduced relative to the validated case. 

For the angular momentum loss rates, we find $1.1\times 10^{29}$~erg and
$1.3\times 10^{29}$~erg, respectively. We also compare the solar
magnetogram results to the solar analogue results. In doing so, we find that the loss rates of both solar cases follow the trends established by the solar analogues. For the planetary magnetosphere sizes seen in Figure~\ref{fig:Rmp}, the high resolution case yields slightly larger radii owing to smaller wind ram pressures. Both solar results follow the analogue trend, with slower stellar winds generally yielding larger magnetospheric radii. \\



\section{DISCUSSION}
\label{sec:Discussion}

We begin this section with a detailed look at the spin evolution of our sample and cool stars in general in Section \ref{sec:spin evolution}, where we also present our relation of spin evolution with time. We continue by discussing the relation between magnetic activity and the Alfv{\'e}n Surface in Section \ref{sec:Alfven}, followed by mass loss rate in Section \ref{sec:massloss}. We then compare our simulation to more observation based studies in Section \ref{sec:observations}. We present a relation for ram pressure evolution with time in Section \ref{sec:magnetosphere}, and address the uniqueness of $\tau$~Bo{\"o}tis in Section \ref{sec:Taudiscuss}. Finally we examine the results of the solar magnetograms in context with the analogues and compared to other studies in Section \ref{sec:soldiscuss}. 

\subsection{Rotation Period and Spin Evolution}
\label{sec:spin evolution}
Within the general paradigm of stellar rotation evolution in which stars spin down with age as a result of angular momentum loss through magnetized winds, younger stars have faster rotation
rates and a larger spread in rotation periods \citep{Soderblom.etal:93, Queloz.etal:98, Bouvier.etal:13, Gallet.Bouvier:13, Johnstone.etal:15, Gallet.Bouvier:15}, 
while older stars (around 0.3~Gyr for solar mass stars \citealt{Gallet.Bouvier:15}) tend to follow the Skumanich Law $\Omega \sim t^{-1/2}$
(\citealt{Skumanich:72}; see however, evidence for departures from this behavior presented by \citealt{van_Saders.etal:16}). 

Once they have shed the their natal circumstellar disks that are thought to modulate rotation through ``disk locking'', stars subsequently spin-up by a factor of 5-10 during their first tens of Myr \citep[e.g.][]{Matt.etal:15, Gallet.Bouvier:15} as a result of the contraction and moment of inertia change that occurs during evolution to the zero-age main-sequence. The stars from our sample appear to follow this paradigm of spin evolution (see Figure \ref{fig:agerot}). The rotation periods of the youngest stars in our sample are then a product of the longevity of the disk-locking phase that can negate the reduction in moment of inertia while it lasts, as well as the subsequent post-disk contraction.
AB Dor is a young (75~Myr) zero-age main-sequence star that has undergone the spin up stage but has yet to spin down significantly.  Since HD~29615 and HD~35296 are only 30 and 35 Myr old respectively, it is possible that they may still be in the spin-up stage, although their longer rotation periods relative to that of AB Dor could also have resulted from a longer disk-locking phase that stifled early spin-up.


It is now widely accepted that rotation, combined with convection, powers the
magnetic dynamo, which is the driver of stellar 
magnetic activity \citep[e.g.][]{Pallavicini.etal:81,Noyes.etal:84,Moss:86,Wright.etal:11}. Thus it is expected that the spin evolution of solar like stars can be used to probe the evolution of the dynamo. As the spin evolution of the star is directly related to angular momentum loss (see equation~\ref{eqn:spindown}), it is straightforward to examine the spin evolution of our sample. Using a power law fit to the angular momentum loss rate and then integrating (see Equation \ref{eqn:spindown}), we find that the spin evolution goes as $\Omega\propto t^{-0.35}$, and when the apparent outlier $\tau$~Boo is removed, as $\Omega\propto t^{-0.51}$--- remarkably close to the Skumanich Law $\Omega\propto t^{-0.5}$. Considering that this law is only applicable for older stars \citep{Skumanich:72}, and that we use a relatively small sample, it is unsurprising that the exact relation is not recovered, but encouraging that a close fit is found. 
This provides some support that our MHD model properly reproduces the physics driving the dynamo powered stellar winds. 

\subsection{Alfv{\'e}n Surface and Magnetic Activity}
\label{sec:Alfven}
\begin{figure*}
\center
 \includegraphics[width=0.4\textwidth]{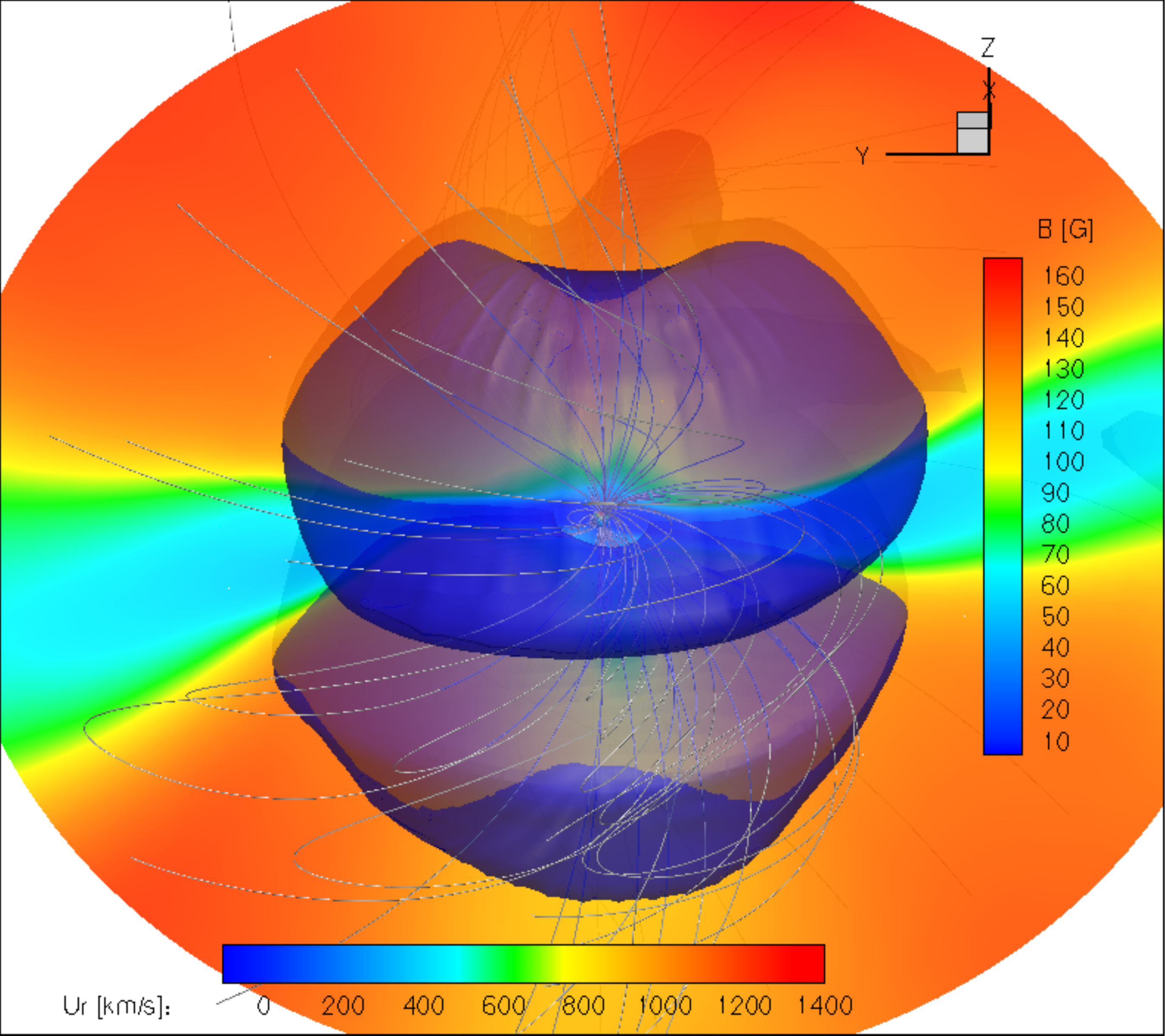}
 \includegraphics[width=0.4\textwidth]{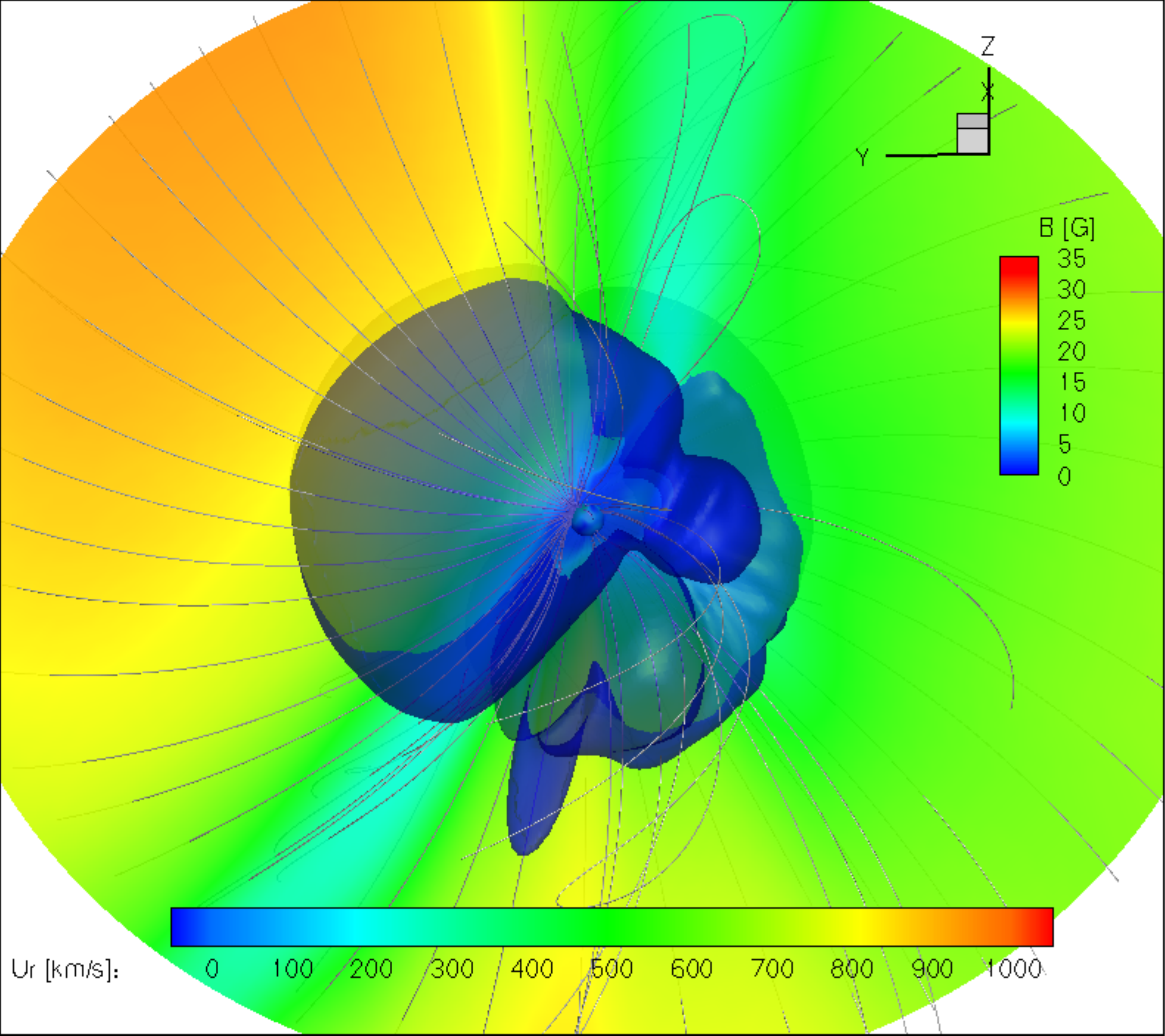}\\
 \includegraphics[width=0.4\textwidth]{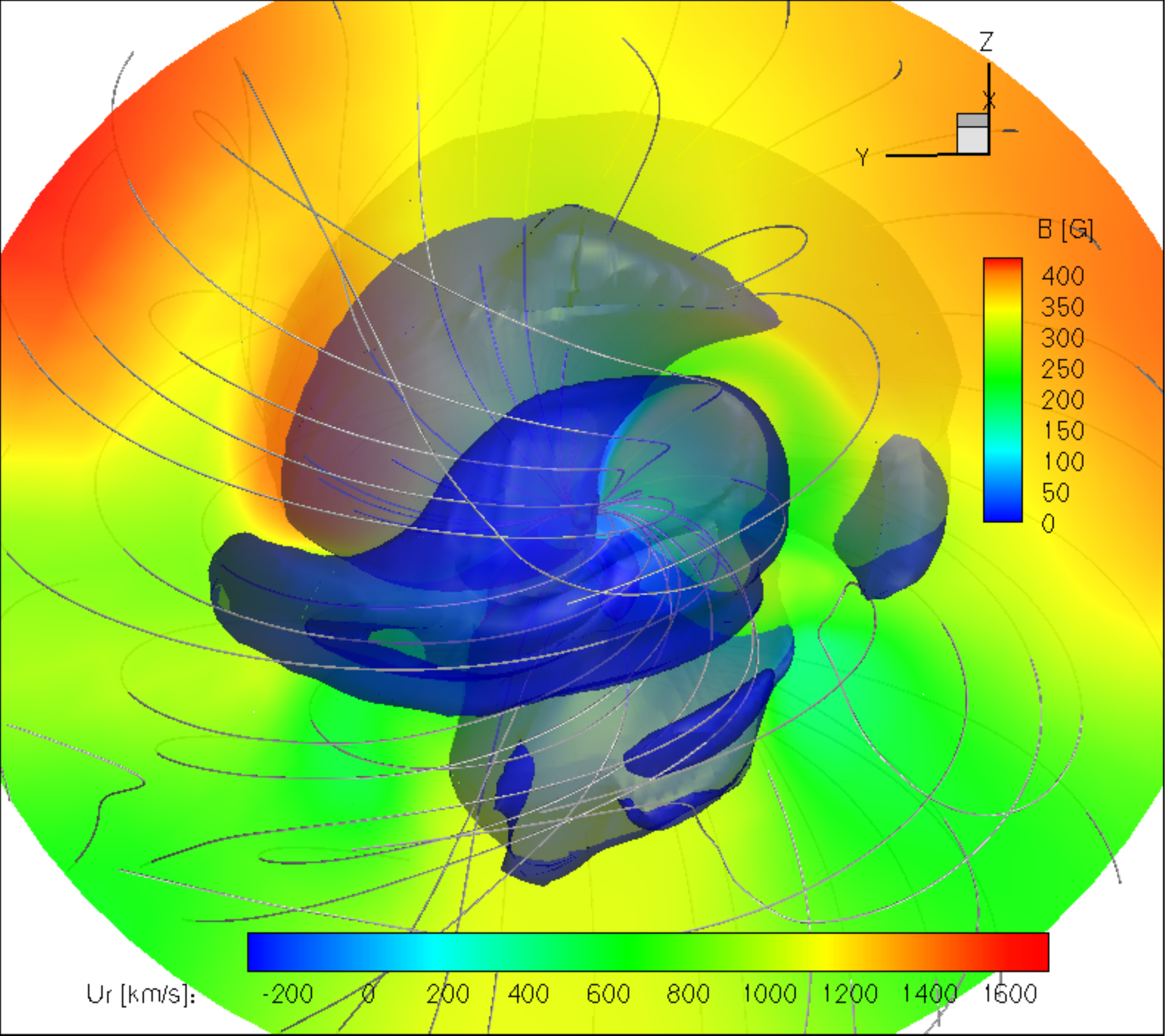}
 \includegraphics[width=0.4\textwidth]{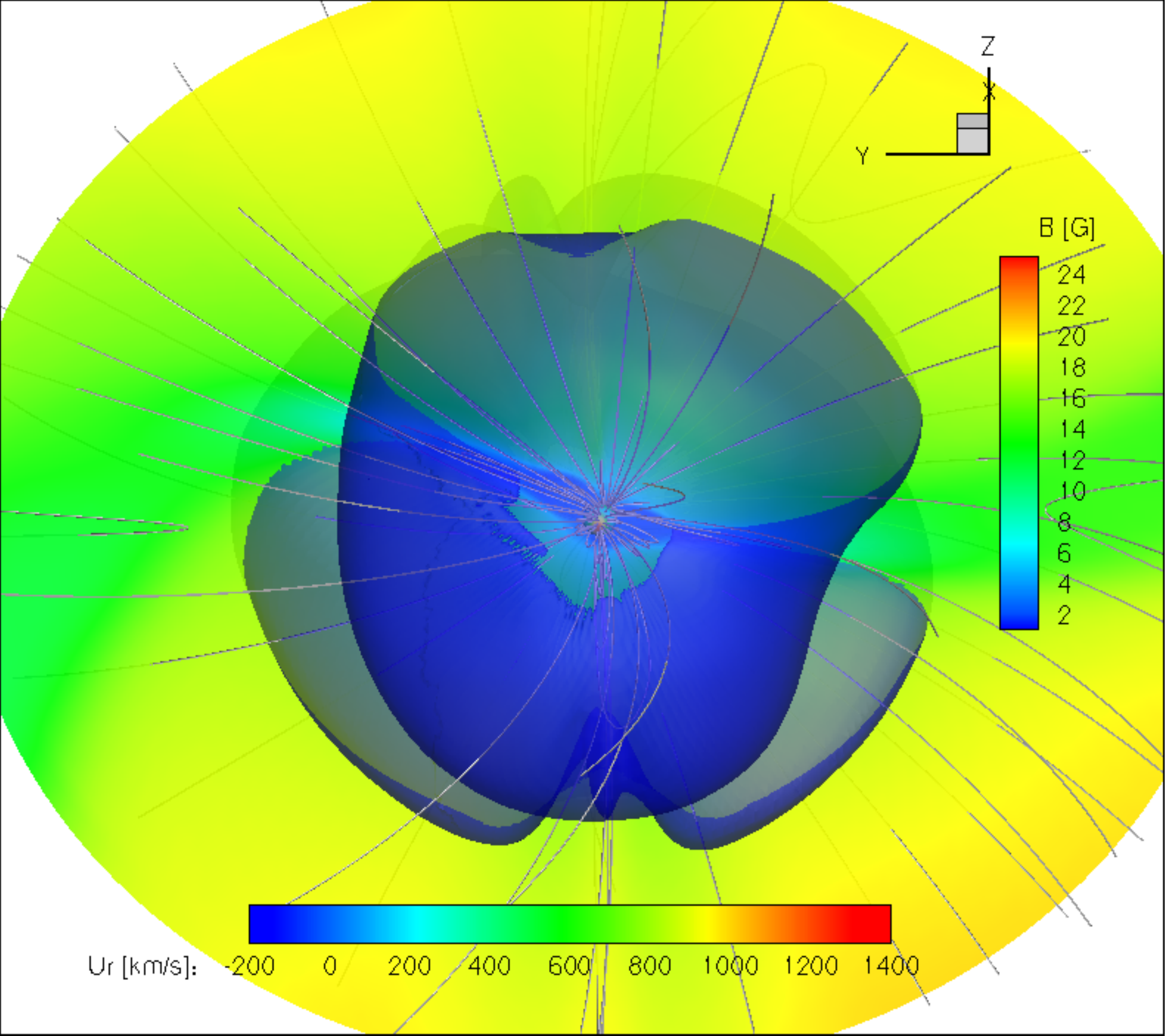}\\
 \includegraphics[width=0.4\textwidth]{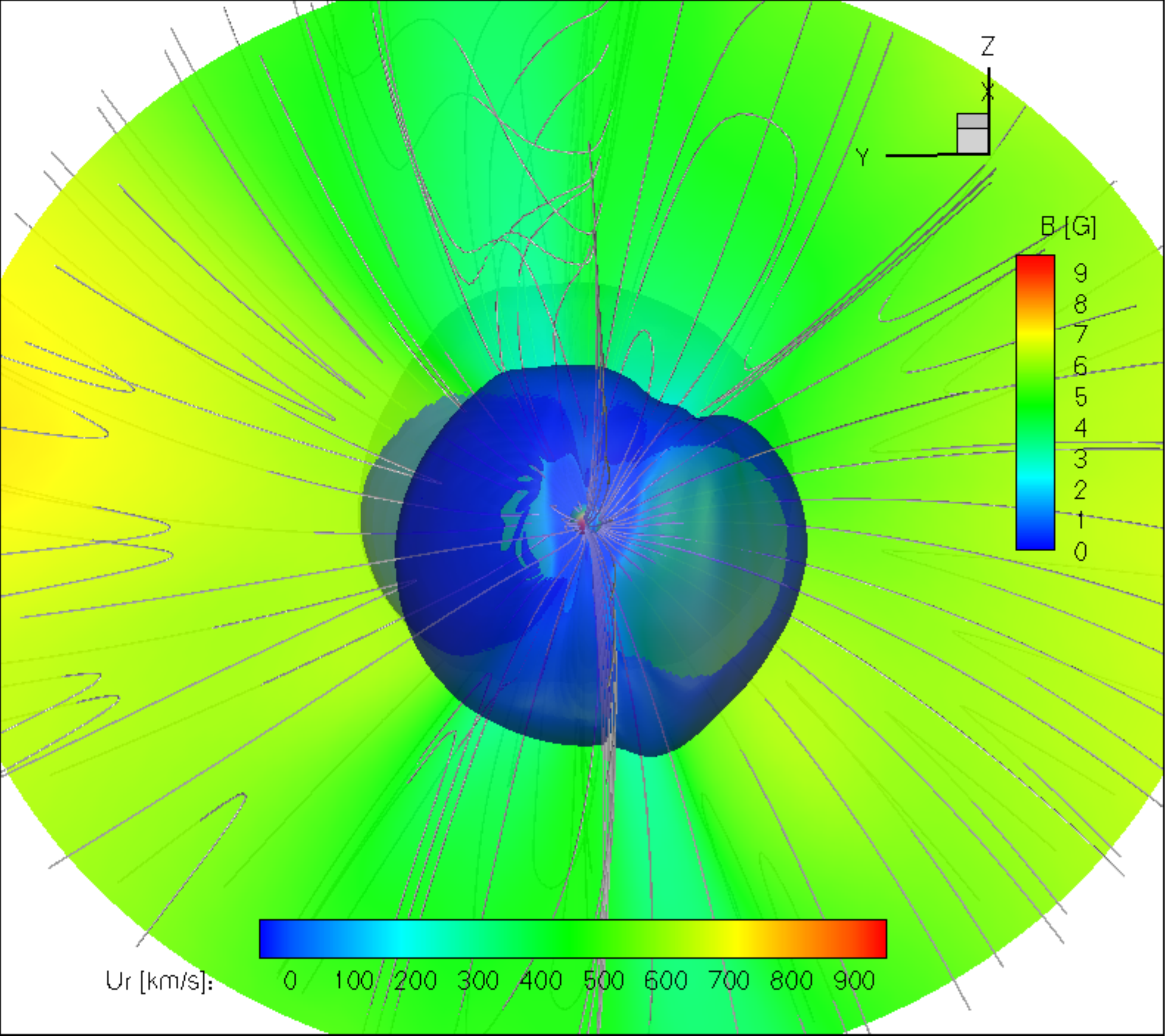}
 \includegraphics[width=0.4\textwidth]{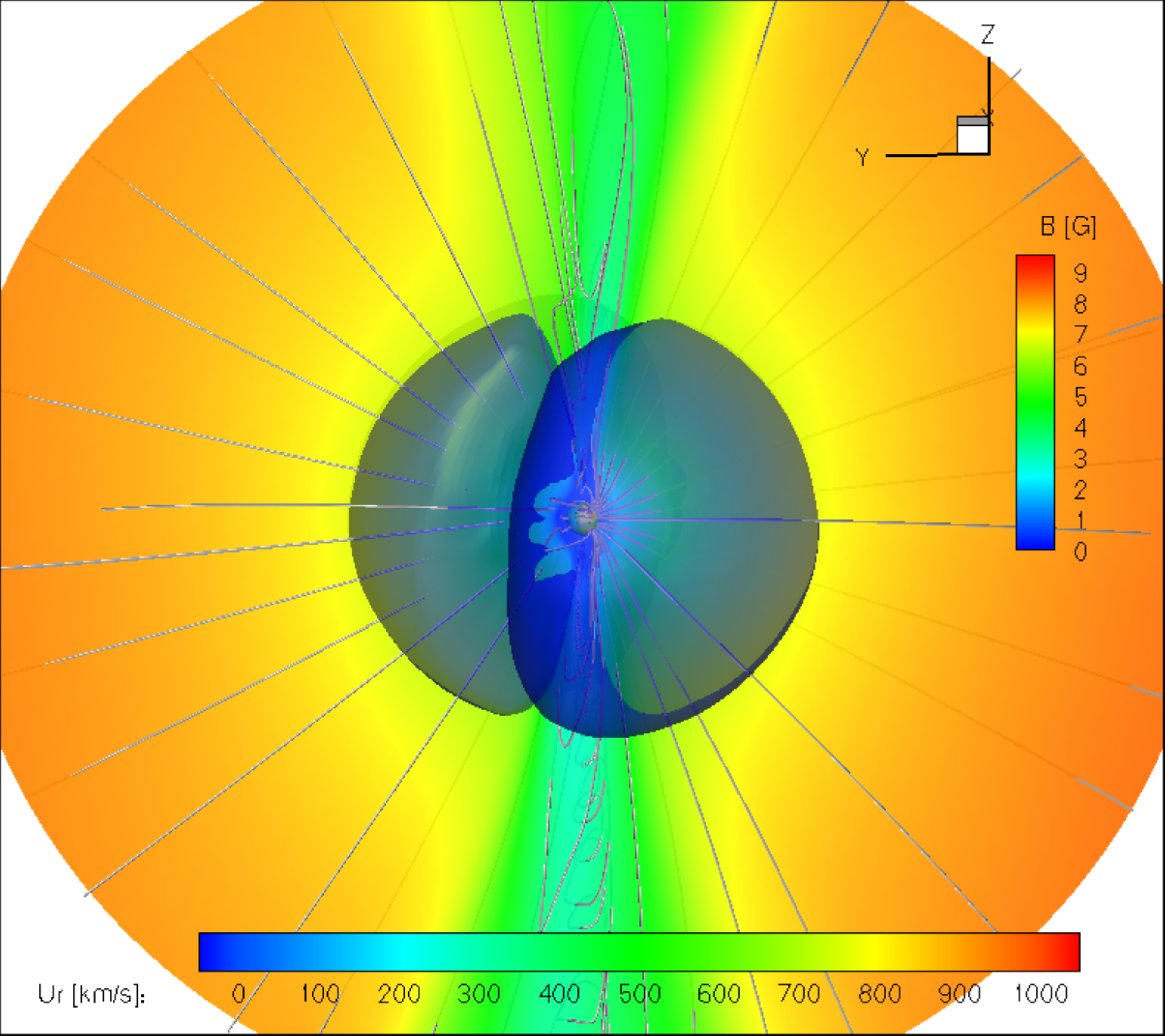}\\
 \caption{Three-dimensional visualisations of the first six simulation
  solutions, in order of ascending age from top left to bottom right: HD 29615, HD 35296, AB Dor, HD
  206860, HD73350 and HD 73256.
  The two-dimensional contour pattern represents a slice
  from the radial wind velocity. The streamlines represent the
  magnetic field lines. The translucent surface is the Alfv\'en
  surface. The sphere at the center of the plot is the surface of the
  star and is flooded to show magnetic field strength. In cases where
  the surface is hard to see, the legend for magnetic field strength
  is still included to give an idea of the range of magnetic field
  strength involved.}
 \label{fig:3dvisa}
\end{figure*}

\begin{figure*}
\center
\includegraphics[width=0.4\textwidth]{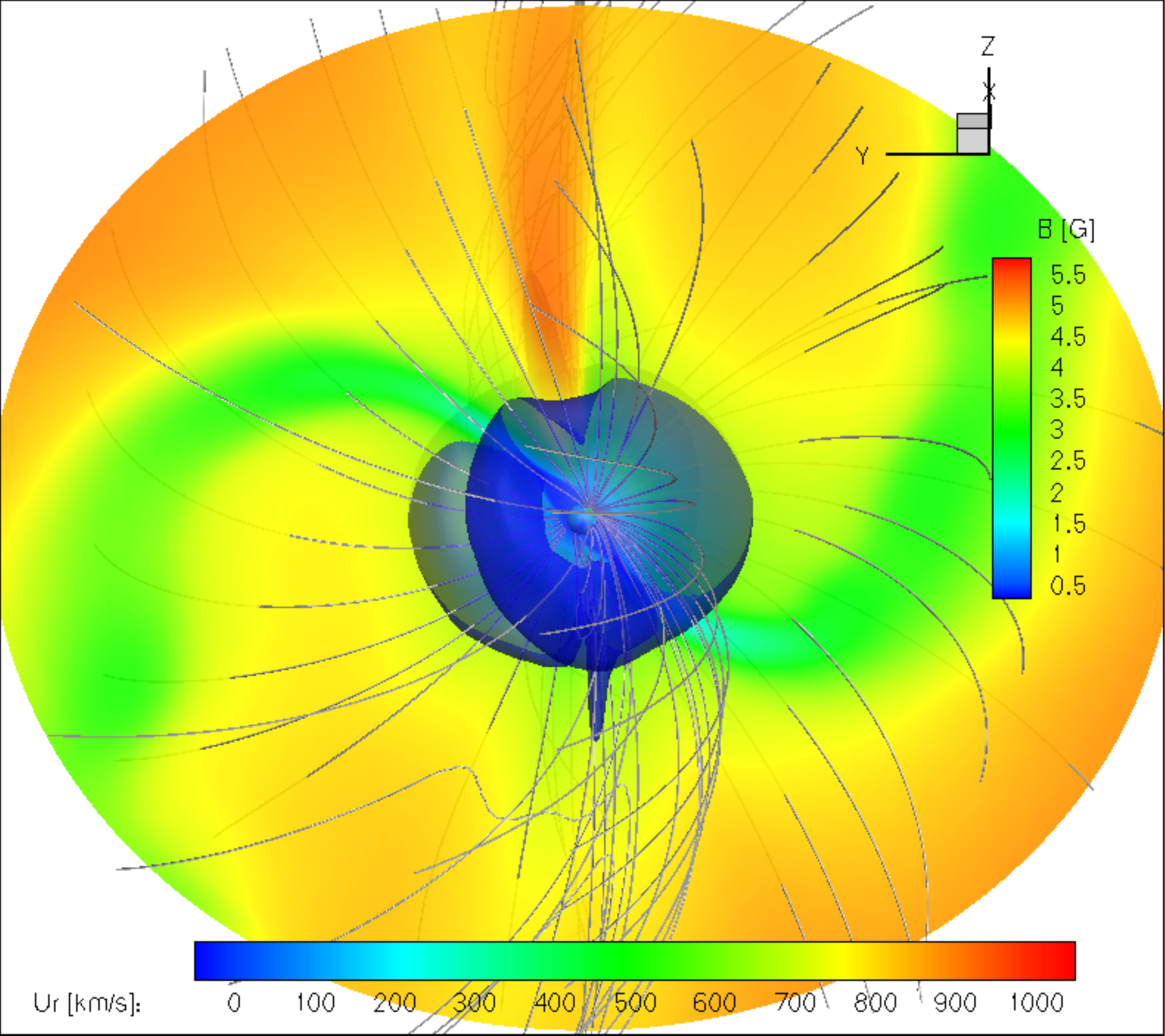}
\includegraphics[width=0.4\textwidth]{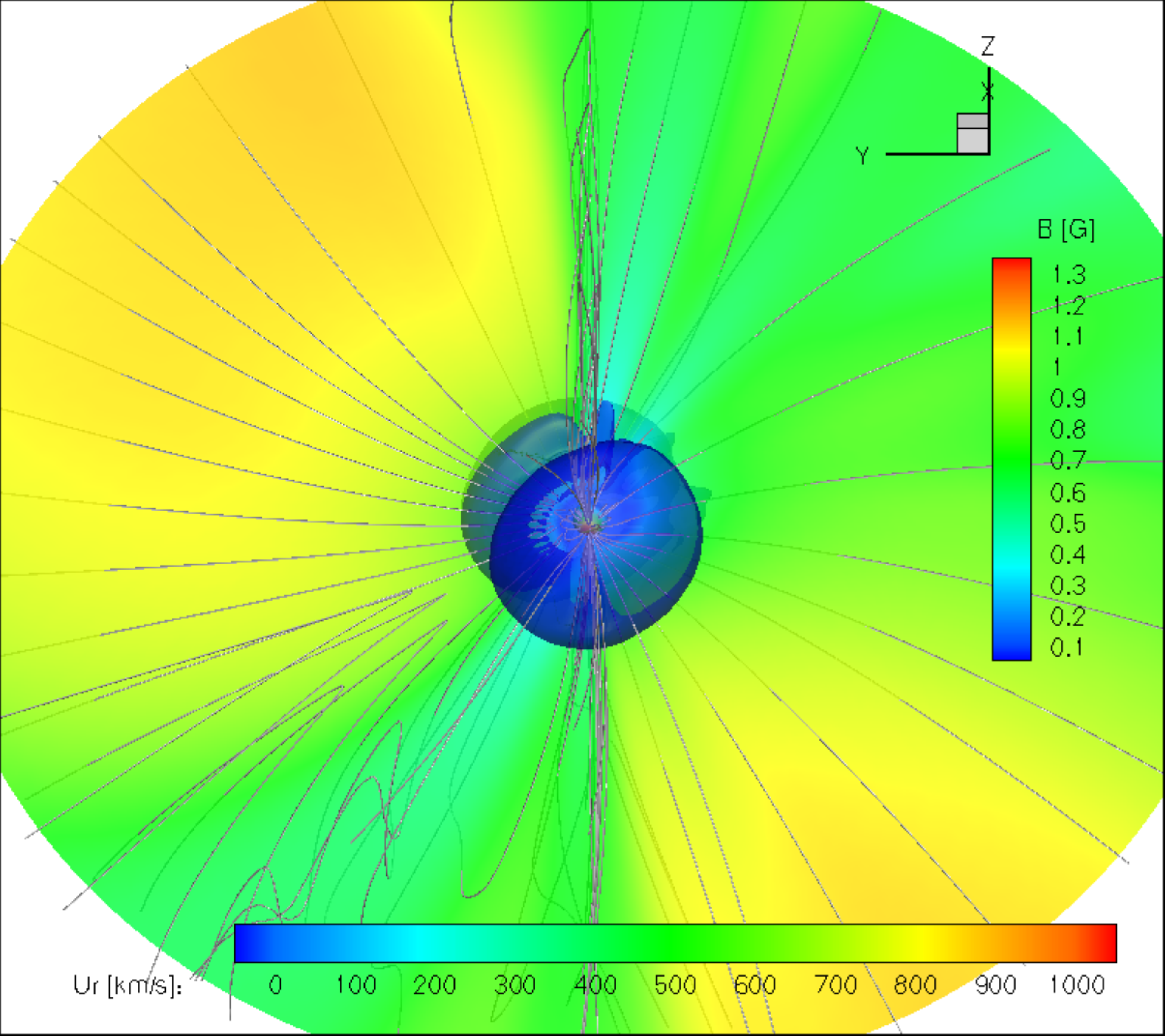}\\
\vspace{0.1in}
\includegraphics[width=0.4\textwidth]{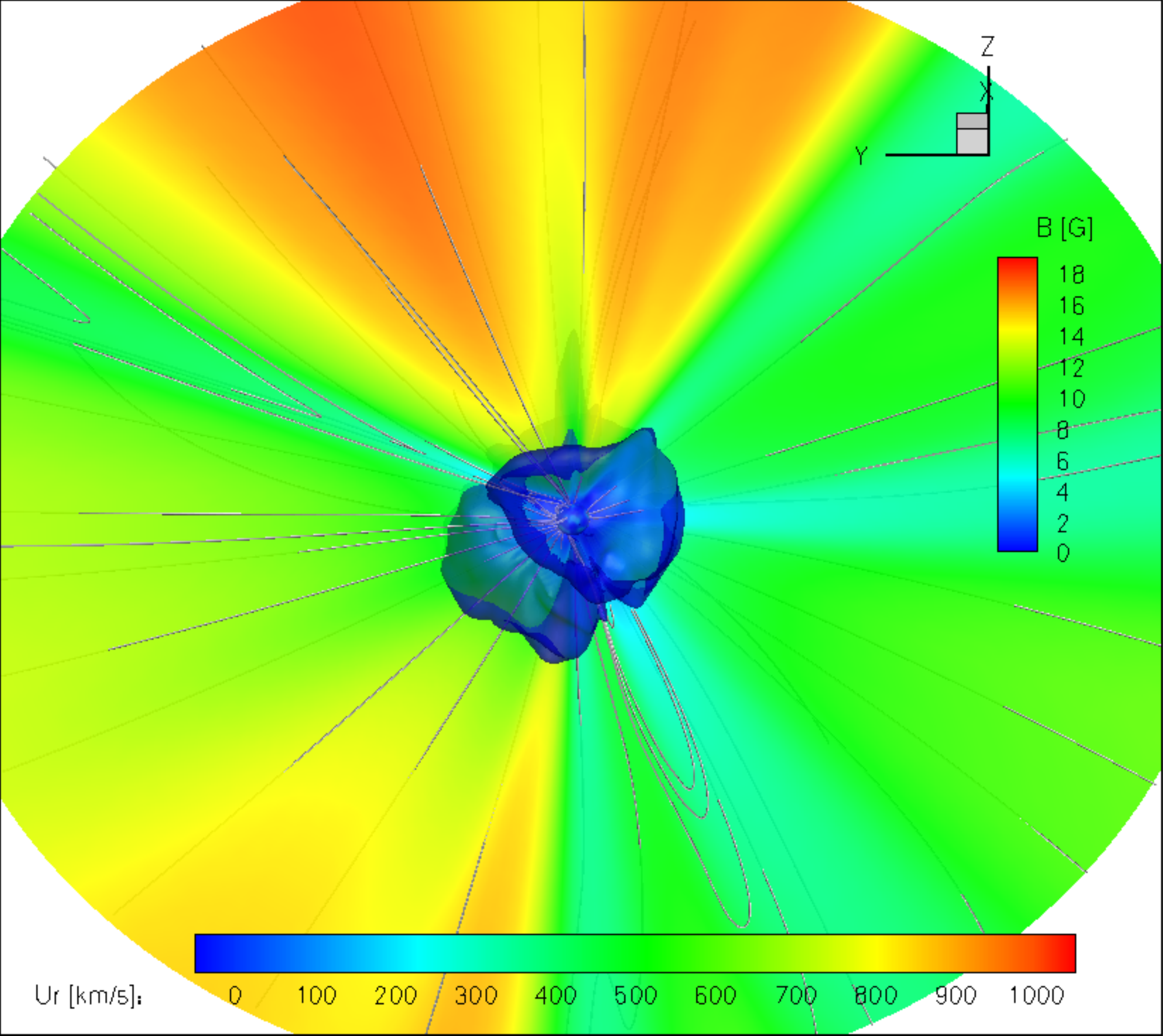}
\includegraphics[width=0.4\textwidth]{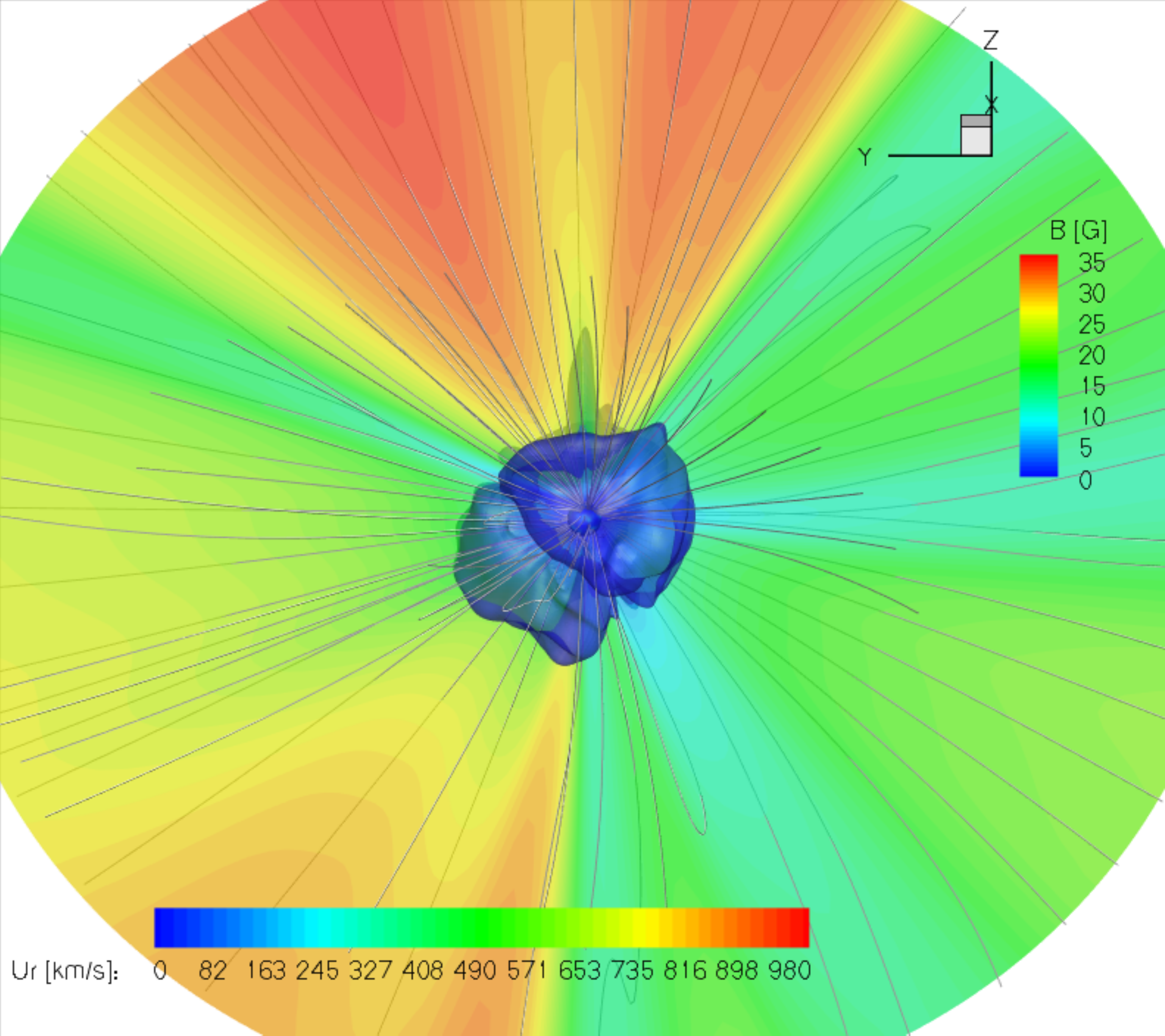}\\
\caption{Three-dimensional visualisations of the four last simulation
  solutions in order of ascending age from top left to bottom right: Tau Boo, HD 76151, high
  resolution solar, low resolution solar.
  The two-dimensional contour pattern represents a slice
  from the radial wind velocity. The streamlines represent the
  magnetic field lines. The translucent surface is the Alfv\'en
  surface. The sphere at the center of the plot is the surface of the
  star and is flooded to show magnetic field strength. In cases where
  the surface is hard to see, the legend for magnetic field strength
  is still included to give an idea of the range of magnetic field
  strength involved.}
\label{fig:3dvisb}
\end{figure*}

Magnetic activity is also naturally linked to the overall strength of the magnetic
field, which is ultimately responsible for 
driving the stellar
wind. The stellar wind environment and size of the Alfv{\'e}n Surface then depend on both the magnetic field strength and rotation period of the stars somewhat degenerately. The Alfv{\'e}n Surface is expected to be larger for stronger fields and faster rotators \citep{Matt.etal:12,Cohen.Drake:14} (see Figures ~\ref{fig:3dvisa} and ~\ref{fig:3dvisb} for visualisation). Several studies have been conducted showing that magnetic field strength has a stronger effect on the Alfv{\'e}n radius, and thus angular momentum loss, than rotation rate, especially for dipolar fields \citep[e.g][]{Pinto.etal:11,Matt.etal:12,Cohen.Drake:14}. 

\cite{Cohen.Drake:14} conducted a detailed analysis of the dependence of the Alfv{\'e}n Surface, mass loss and angular momentum loss on magnetic dipole component, rotation period and base density using an MHD simulation grid. They found that the Alfv{\'e}n Surface increases its overall size with both magnetic field and rotation period, and that mass loss is greater for stronger fields, which have slower but denser stellar winds. While our results agree that stronger magnetic fields yield larger mass loss rates, we find that the winds for these strong fields are faster than for the weak fields (see Figures \ref{fig:3dvisa} and \ref{fig:3dvisb}). We believe these qualitative discrepancies are due to the fact that our analogue magnetograms do not exhibit dipole-like behaviour (see Figure~\ref{fig:magnetograms}). Many studies have been conducted to show that the effect of the magnetic field is altered and reduced for more complex morphologies \citep[see][for example]{Pinto.etal:11,Cohen.Drake:14, Garraffo.etal:15} though it remains important as seen in our results for AB Doradus and HD~29615, which both have strong magnetic fields and Alfv{\'e}n Surfaces that extend almost up to 80 and 60 stellar
radii respectively. 

As rotation is believed to be linked to the magnetic dynamo and thus the magnetic field, the role rotation plays for the angular momentum and mass loss rates of fast rotating young stars is of some interest. \cite{Airapetian.Usmanov:16} find from 3D MHD simulations that for rotation periods of 2.5-25 days, rotational effects such as centrifugal force are negligible for mass loss rate and wind speed such that we can conclude that rotation period dependencies in our results are due to its influence on the attendant magnetic field. Stronger magnetic fields in the absence of other changes in wind parameters will lead to larger Alfv{\'e}n surfaces and so increased angular momentum loss rates.  Greater Poynting flux from stronger magnetic field is also expected to drive more mass loss, which acts to decrease the Alfv{\'e}n surface size.  The interplay between the two effects can be complex---see, e.g., the discussion of \citet{Matt.etal:12} for the case of Parker-type winds.

\subsection{Mass Loss Rate}
\label{sec:massloss}
We see from our results (see Figures \ref{fig:Bavg} and \ref{fig:rot}) that the mass loss rate is anticorrelated with rotation period. In order to compare stars of different spectral type, rotation is often expressed in terms of the Rossby number, defined as $Ro=P_{rot}/\tau_c$, where $P_{rot}$ is the rotation period and $\tau_c$ is the convective turnover time. Recently, a study conducted by \cite{See.etal:17} used a potential field source-surface model to approximate MHD calculations of angular momentum and mass loss rates on 66 ZDI mapped solar analogues as a function of Rossby number. They find that the loss rates decrease with Rossby number as expected, with angular momentum rates similar to our results, but with mass loss rates which are acknowledged as being too high. Aside from the general trends expected, \citet{See.etal:17} also make note of a saturated loss rate regime for low Rossby numbers ($Ro<0.1$; see also \citet{Ardestani.etal:17}) that coincides with the general magnetic saturation seen through the magnetic activity diagnostic of X-ray emission \citep[e.g.][]{Wright.etal:11}.  \citet{Wright.etal:11} found saturation limits in terms of both an empirical Rossby number as well as rotation period; the latter as a function of stellar mass.  For solar mass stars, the convective turnover time is about 20 days, and saturation occurs at a rotation period of 2.6 days.  Of our sample, AB~Dor is clearly in the saturated regime, but the remainder are either at the saturated-unsaturated limit (HD~29615) or else are unsaturated. It would be interesting in future studies to include faster rotators to probe the early spin-down phase of solar mass stars and explore further the magnetically saturated wind regime and mass loss saturation effects found by \cite{See.etal:17}.

The mass loss rates found in this study are between $10^{-12}$ -- $10^{-15} M_{\odot}/yr$, an order of magnitude lower than the range presented in most studies \citep[e.g][]{Cranmer.Saar:11,Airapetian.Usmanov:16,See.etal:17}. This potential discrepancy is explained due to our omission of magnetic field scaling of the ZDI magnetograms, as explained in Section \ref{sec:solresults}. When looking at the mass loss rates as a function of age, it is interesting to compare our results to those found in other MHD simulations such as \citet{Airapetian.Usmanov:16}, and those found in theoretical models such as that of \citet{Cranmer.Saar:11}. It is reassuring that the results of \citet{Airapetian.Usmanov:16} agree with our simulation results, excepting the order of magnitude for mass loss rates. Furthermore, these values are also consistent with the study of \citet{Wood.etal:05}, which were based on observations of Ly$\alpha$ absorption in stellar ``astrospheres'' (see Section \ref{sec:observations}). However, \citet{Airapetian.Usmanov:16} only take their simulations back to 0.7Gyr, whereas \citet{Cranmer.Saar:11} use a physically motivated model for the winds of cool stars, and cover a range of ages down to 1Myr. The models presented in that work follow turbulent MHD motion from stellar convective zones driven by gas pressure for solar like stars relevant to our study. \citet{Cranmer.Saar:11} present a mass loss rate model as a function of age covering the entire range of our sample, which is very consistent with the trends we see from our mass loss rates, including the young "spin-up" phase for stars between 10 -- 50 Myr.

We can also compare our simulation results to semi-empirical works. \citet{Johnstone.etal:15} studied the wind and mass loss evolution of sun-like stars, while \citet{Ardestani.etal:17} adopted a semi-empirical model approach, both in an attempt to build a general spin down model for stars of ages ranging from 10~Myr to 4~Gyr. Both use the wind torque relation derived in \citet{Matt.etal:12}, finding a relation for mass loss rate to stellar mass, radius and rotation, though the semi-empirical model also factors in mixing time for angular momentum transfer inside the stellar envelope. Comparing their qualitative results of mass loss evolution to the trends found from our study, we again see the same general evolution that mass loss rate decreases with age and as rotation period increases. However, they find a steeper slope for these relations than our results, with $dM/dt \propto P^{-1.33}$ \citep{Johnstone.etal:15}, and $dM/dt \propto P^{-1.3}$ \citep{Ardestani.etal:17}, where P is rotation period, while we find $dM/dt \propto P^{-0.90}$. 

\subsection{X-ray and Indirect Mass Loss Rate Observations}
\label{sec:observations}
X-ray and extreme ultraviolet (EUV) emission are often used to probe stellar magnetic activity \citep{Gudel.etal:96,Ribas.etal:05,Guinan.Engle:09}, with X-ray saturation being of particularly interest \citep{Wright.etal:11,See.etal:17}. From our results in Figure \ref{fig:Lx}, it is difficult to see a saturated X-ray regime due to the small sample. However, we clearly see that the mass and angular momentum loss rates appear closely correlated to the X-ray flux. This is due to the fact that all these quantities are indicators of the magnetic dynamo's activity. This does not indicate that one is the causation of the others. For example, \citet{Cohen:11} has found that the Sun's mass loss rate stays at an average of $2\times 10^{-14} M_{\odot}/yr$ without following the variations of solar X-ray flux. This is explained by the mass loss rate begin heavily dependent on stable open magnetic flux, while X-ray emission also depends on the more variable closed magnetic flux. Since X-ray flux correlates well with magnetic activity, it also correlates very well with Rossby number, defined as rotation period divided by convective turnover time $\tau$. Thus if $\tau$ is known for a certain spectral type, rotation period can be inferred, albeit with some uncertainty, from X-ray luminosity, and using gyrochronology for stars older than $\sim$1Gyr, age can be inferred. However, this method has some flaws which have been detailed, notably that this method is difficult to apply for younger stars, and that there exists a large spread of ages for a certain rotation period \citep{Barnes:03,Barnes.Kim:10}. Since we already take into account uncertainty in our sample's ages and rotation periods, and there are also some disagreements for the spectral types in literature, we omit any derivation our sample's ages using X-ray flux. 

\citet{Wood.etal:05,Wood.etal:14,Wood.etal:15} indirectly measured the mass loss rates for cool main sequence stars using absorption in high dispersion spectra of Ly$\alpha$~1215.7 Angstrom in stellar astrospheres. The general trends observed here are consistent with those found in this work, notably that mass loss rate decreases with age and the values sit in an interval between $10^{-12}$ to $10^{-14}$ solar masses per year. However, \citet{Wood.etal:15} suggest that young stars with ages less than 0.7Gyr may in fact have weaker winds more akin to those of the Sun, rather than the powerful winds inferred from the high mass loss rates in this work. The star $\pi^1$ UMa is taken as an example. This is a G1.5 spectral type star, with an age of $\sim$300Myr and rotation period of $\sim$5 days, similar to HD 206860 in our sample. While we find a mass loss rate for HD 206860 an order of magnitude higher than for the Sun, \citet{Wood.etal:15} find a value for $\pi^1$ UMa lower than for the Sun. This may be explained in terms of the complexity of the magnetic field. 
The $\pi^1$ UMa detection is very interesting in the context of the current discussion on young stars' spin-down. There is increasing evidence indicating that young, active stars can store larger fractions of their magnetic flux in higher order spherical harmonics \citep{Donati.etal:08, Donati.Landstreet:09} which they lose as they age.  This complexity should lead to a suppression of dM/dt and angular momentum loss \citep{Reville.etal:15a,Reville.etal:15b,Garraffo.etal:15,Garraffo.etal:16}. These results combined may indicate that very active stars suffer a suppression of mass loss rates due to magnetic field complexity. The bimodal distribution of rotation periods in open cluster observations suggest that for any age, there are some stars that behave like dipoles and some that behave like more complex morphologies not losing angular momentum or mass at an efficient rate. HD 206860 does not show a very complex field, either due to the lack of sensitivity of the ZDI technique or perhaps this star's complexity has already evolved towards a simpler one. 
Such a dispersion in dM/dt values might then be naturally expected for young stars.

In summary, the overall trends here established by the 
solar analogue sample are as expected and generally agree with previous studies.

\subsection{Planetary Magnetosphere Evolution}
\label{sec:magnetosphere}

The results obtained for the range of planetary magnetosphere
sizes also confirm the expected trend of an increase in size as stars
age, since weaker winds will exert less ram pressure. The size of the magnetosphere depends on the pressure exerted by the stellar wind, which in turn is related on the velocity and density of the wind. When probing the densities and velocities of the stellar winds, we find that the slower winds consistently have a higher density. A switch occurs in which winds density is a more dominant factor in the ram pressure than velocity for younger stars, while the opposite is true for older stars. 

\begin{figure*}
\center
\includegraphics[trim=0.1in 0.1in
  0.1in 0.1in,clip,width=0.8\textwidth]{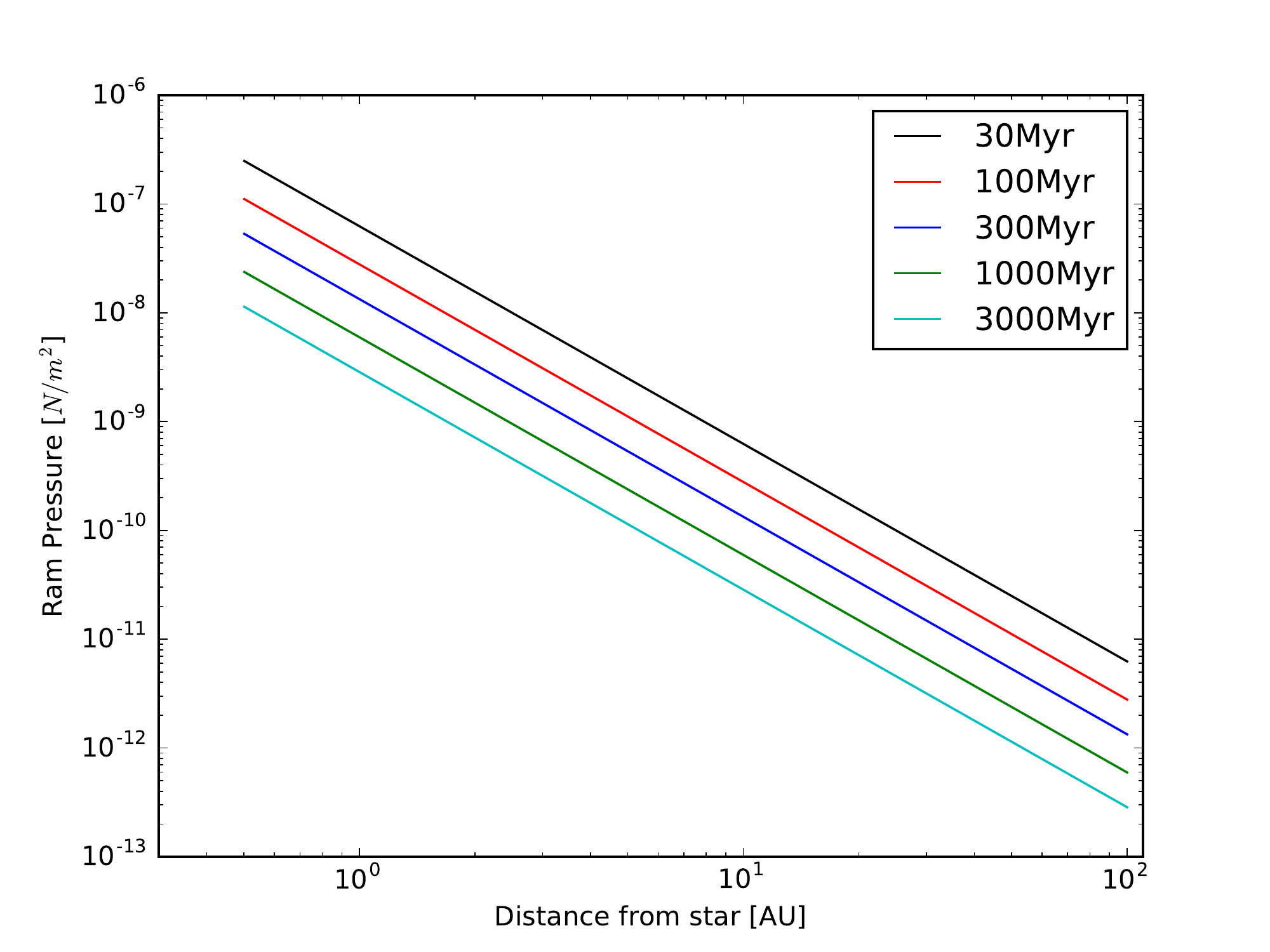}
\caption{Plot of ram pressure with distance at loci of different ages.} 
\label{fig:Rampress}
\end{figure*}

Using the wind velocity and density results, we fit a power law relation to the mean of the minimum and maximum ram pressures listed in Table~\ref{tab:magnetograms} as a function of age. By scaling this relation with orbital distance, we can then derive a relationship that describes the evolution of the wind ram pressure throughout the solar system as a function of time: 

\begin{equation}
 \overline{P}_{ram} = 6.10\times10^{-7}~t^{-0.67}/r^2 
 \label{eqn:rampress}
\end{equation}

where t is age in Myr and r is distance in Au. This is illustrated as a set of loci corresponding to the pressure as a function of orbital radius for different ages in Figure~\ref{fig:Rampress}. As expected, the ram pressure decreases with age, consistent with the star spinning down and the stellar winds undergoing a steady secular loss in power. This is in agreement with previous studies that have found that the solar wind pressure at Earth's magnetopause was much greater in the past, potentially up to two orders of magnitude between present day Gyr ages and early $\sim$ 10 Myr ages \citep{Sterenborg.etal:11,Airapetian.Usmanov:16}.

\subsection{Tau Bo{\"o}tis as a Potential Outlier}
\label{sec:Taudiscuss}
Of the eight solar analogues examined in this study, $\tau$~Bo{\"o}tis stands
out the most as a potential outlier. In Figure ~\ref{fig:agerot}, we
see that $\tau$~Bo{\"o}tis rotates at a much faster rate than any
of the other stars of comparable age, with a rotation period of merely
3 days. Since it is generally accepted
that rotation is responsible for powering the magnetic dynamo, one would expect the average surface magnetic field strength to be more akin to those of the young, fast rotators. Figure \ref{fig:Bavg} clearly shows that this is not the case, and indeed $\tau$~Bo{\"o}tis fits very well with stars of similar ages, even when they are rotating much slower. Indeed, its mean surface magnetic field is found to be 1.26~G, while comparable rotators HD 29615 and HD 35296
with 2.32 and 3.9 day periods both have much larger
average magnetic fields of 119~G and 68.5~G
respectively. As such, with a rotation period of 3 days, one could
expect to have an average field strength somewhere
between these two values. 

$\tau$~Bo{\"o}tis' angular momentum loss rate places it comfortably with the young stars, but its mass loss rate places it more with the old stars of our sample (see Figures~\ref{fig:age} and \ref{fig:rot}). Since angular momentum loss rate is directly proportional to angular frequency (see equation~\ref{eqn:angmomloss}), it is unsurprising that $\tau$~Bo{\"o}tis fits well with other young rotators. However, mass loss rate (see equation~\ref{eqn:massloss}) depends only on the wind velocity at the Alfv{\'e}n Surface and the wind density. Furthermore, as discussed in Section~\ref{sec:spin evolution}, when removing $\tau$~Bo{\"o}tis from the fit shown in Figure~\ref{fig:age}, we find a relation $\Omega \sim t^{-0.51}$ (shown in red in the bottom panel), which is even closer to the Skumanich Law than when $\tau$~Bo{\"o}tis is included in the fit (which yields $\Omega \sim t^{-0.36}$). While this should not be used as support for treating this particular star as an outlier, that the spin-down relation moves in the direction toward that observed is encouraging.

There are several reasons that could seek to explain why the $\tau$~Bo{\"o}tis results appear to be so out of place. As mentioned in Section \ref{sec:TauBoo}, $\tau$~Bo{\"o}tis is host to a hot Jupiter with an orbital period of 3.3 days with potentially strong tidal effects \citep{Butler.etal:97,Catala.etal:07,Donati.etal:08, Fares.etal:09}. The magnetic field of the giant planet may also interfere with that of the star, further altering the results derived from our simulations. Furthermore, there are large uncertainties on the age of $\tau$~Bo{\"o}tis, with values generally ranging between 1.3-2.1 Gyr (see Section \ref{sec:TauBoo}).Thus it is possible that $\tau$~Bo{\"o}tis is in fact a true outlier for physical reasons. 

There may also be technical reasons behind the apparent uniqueness of the $\tau$~Bo{\"o}tis results. It was noted during its spectropolarimetric observations that "the Zeeman signatures of $\tau$~Boo are extremely small,"~\citep{Fares.etal:09} which could account for the weak field magnitude relative to its rotation rate. Furthermore, we see in Figure~\ref{fig:magnetograms} that the map for $\tau$~Bo{\"o}tis appears to be of poor quality relative to the others and it is possible that the ZDI-derived magnetogram is flawed. We also note that \citet{Vidotto.etal:12} conducted a similar wind model study on $\tau$~Bo{\"o}tis and found loss rates of $2.7\times10^{-12}M_{sol}$ per year and $1.5\times10^{32}erg$. These results are 2 and 1 orders of magnitude above our already high results, though are based on less physically realistic Parker-type winds.  

In summary, until additional ZDI observations of $\tau$~Bo{\"o}tis are acquired and the present results can be verified they should be treated with caution. 


\subsection{Solar Results in Context}
\label{sec:soldiscuss}

In order to investigate how the case of the Sun fits in with the results for the solar analogues it is interesting to bring the two
solar magnetogram results into the picture. Figures \ref{fig:Bavg}, \ref{fig:age} and \ref{fig:rot} show
these results as upwards and downwards triangles for the high and low
resolution magnetograms respectively. It is clear that the difference in resolution does not appear to change the results much, as the magnetic field strength and loss rates are essentially identical. As noted earlier, both high and low resolution cases fit the analogue trend. For the mass loss rates, \cite{Sterenborg.etal:11}
reported a rate of $4.28\times10^{-14}$ solar masses per
year for the same high resolution magnetogram, while the solar wind study by
\cite{Cohen:11} found that the average mass loss rate of the Sun is
about $2\times10^{-14}$ solar masses per year.  Our mass loss rate is $5.7\times10^{-15}$ solar masses per year---slightly lower than found in other studies. As mentioned in Section \ref{sec:solresults}, we chose to not apply the 2--3 multiplicative correction to the input magnetograms' magnetic fields, which are needed to obtain the right magnitude of the magnetic flux at 1~AU \citep{Cohen.etal:08,Riley.etal:12,Linker.etal:12}, in order to avoid arbitrary scaling of data. Since we did not rescale any magnetic maps, finding a lower mass loss rate for the Sun relative to the validated case is expected.

The inclusion of two different resolution solar magnetograms in this study also serves to test the accuracy of ZDI maps and to address the question of whether small-scale features such as spots could significantly influence mass and angular momentum loss rates. It has been shown that for an underlying dipolar field morphology, the large scale structure dominates smaller features such as sunspots in controlling wind properties \citep{Garraffo.etal:13}.  
The Sun is not a dipole during solar maxima, thus allowing the influence of small scale features to be tested. 
We find that the difference in resolution between the MDI and WSO observations is insignificant for the purposes of predicting the wind global properties.

Several previous studies have simplified the problem of magnetic field morphology by assuming purely dipolar fields when simulating the younger Sun \citep[e.g.][]{Matt.etal:12, Cohen.Drake:14, Matt.etal:15, Airapetian.Usmanov:16}. It is not known what the underlying magnetic morphology of solar analogues is in general and the magnetograms in Figure~\ref{fig:magnetograms} are not
clearly indicative of dipoles. Field morphology also likely changes through the lifetime of the star \citep[e.g.][]{Garraffo.etal:15}.
This must be kept in mind when placing this work in context with other studies, especially since dipolar fields were shown by \citep[e.g.][]{Garraffo.etal:15} to be more efficient at removing mass and angular momentum from the star than multipole fields. Furthermore, ZDI is easiest to apply to bright, nearby, fast-rotating stars exhibiting both stronger magnetic field and Doppler signatures 
\citep{Donati.Brown:97}. As such, the younger and faster rotating stars
of our sample are generally expected to have better data quality than the much
older and more slowly rotating stars. This could potentially induce a bias into the results. That this is probably not a serious effect is 
supported by the trends obtained from ZDI magnetic maps by \citet{Vidotto.etal:14a} that agree with expectations of the evolution of magnetic activity with age, and by the fact that 
our solar analogue sample reproduces a Skumanich-like spin-down trend.

Considering the solar analogues truly as proxies for a younger Sun, we find that the general trends established are consistent with previous studies employing many various models for mass loss and spin evolution of the Sun \citep[e.g.][]{Cranmer.Saar:11,Johnstone.etal:15,Airapetian.Usmanov:16,Ardestani.etal:17,See.etal:17}.



\section{CONCLUSIONS}
\label{sec:Conclusions}

A sufficient body of magnetic maps of solar
analogues now exists with which to perform detailed numerical simulations that can be used to investigate the history of the Sun and solar system interplanetary environment.

A state-of-the-art 3-D MHD model of the solar wind based on the dissipation of Alfv\'en wave turbulence has been applied to a sample of eight solar-like stars with a range of ages and activity levels.  The only variable in the model input was the spatially-dependent radial magnetic field at the lower boundary.  Test case solar simulations employed lower resolution WSO and high resolution MDI magnetograms, while stellar simulations used magnetograms derived from Zeeman-Doppler imaging. 

Both angular momentum and mass loss rates derived from the simulations decrease with stellar age, as expected. From the angular momentum loss rates we derived the spin down rate and find that angular velocity decreases with time according to $\Omega \propto t^{-0.36}$.  When omitting one outlier star, the planet hosting $\tau$~Boo, whose magnetogram, magnetic and rotation properties appear anomalous, the spin-down relation is described by $\Omega \propto t^{-0.51}$.  This is remarkably similar to the Skumanich relation $\Omega \propto t^{-0.5}$.

The difference in spatial resolution between MDI and WSO solar magnetograms does not lead to significant differences in the solar wind simulation results. 

The simulation results were used to compute the wind ram pressure for each stellar case.  This was used to investigate the magnetospheric stand-off distance for an Earth like planet with an equatorial magnetic field strength of 0.3~G situated at 1~AU.  At early times of a few tens to a hundred million years, such a magnetosphere would have been compressed to half of its present-day extension.  An expression for the the wind ram pressure as a function of radial distance and time was derived that indicates that solar wind pressure has declined by a factor of about 50 since the Sun reached the zero-age main-sequence.

 With improvements in ZDI
observational methods and in capabilities of the model, future data should allow for the Sun's history to be
simulated even more realistically through the use of Sun-like stars
as solar proxies.\\   

We would like to thank our anonymous referee for a very thorough and constructive review of our work. QP thanks the High Energy Astrophysics Division of the Harvard-Smithsonian Center for Astrophysics for accommodating him on such a project. This work was carried out using the SWMF/BATSRUS tools developed at The University of Michigan Center for Space Environment Modeling (CSEM) and made available through the NASA Community Coordinated Modeling Center (CCMC). The simulations were performed on the SI Hydra cluster. CG acknowledges support from the NASA Living with a Star program grant NNX16AC11G.
JJD was funded by NASA contract NAS8-03060 to the {\it Chandra X-ray Center} and thanks the Director, Belinda Wilkes, for continuing support.



\acknowledgments




\bibliographystyle{aasjournal}

\end{document}